%% file: main.tex
\documentclass[acmsmall]{acmart}

\setcopyright{cc}
\setcctype{by}
\acmDOI{10.1145/3728950}
\acmYear{2025}
\acmJournal{PACMSE}
\acmVolume{2}
\acmNumber{ISSTA}
\acmArticle{ISSTA073}
\acmMonth{7}
\received{2025-02-27}
\received[accepted]{2025-03-31}

\AtBeginDocument{%
  \providecommand\BibTeX{{%
    \normalfont B\kern-0.5em{\scshape i\kern-0.25em b}\kern-0.8em\TeX}}}

\usepackage{lineno}
\usepackage[]{algorithm2e}
\usepackage{algorithmic}
\usepackage{textcomp}
\usepackage{caption}
\usepackage{float}
\usepackage{subcaption}
\usepackage{multirow}
\usepackage{graphicx}
\usepackage{booktabs}
\usepackage{minted}
\usepackage{enumitem}
\usepackage{listings}

\usepackage{colortbl}

\usepackage{todonotes}

\definecolor{codegreen}{rgb}{0,0.6,0}
\definecolor{codegray}{rgb}{0.5,0.5,0.5}
\definecolor{codepurple}{rgb}{0.58,0,0.82}
\definecolor{backcolour}{rgb}{0.95,0.95,0.92}

\lstdefinestyle{mystyle}{
  backgroundcolor=\color{backcolour},   
  commentstyle=\color{codegreen},
  keywordstyle=\color{magenta},
  numberstyle=\tiny\color{codegray},
  stringstyle=\color{codepurple},
  basicstyle=\ttfamily\footnotesize,
  breakatwhitespace=false,         
  breaklines=true,                 
  captionpos=b,                    
  keepspaces=true,                 
  numbers=left,                    
  numbersep=5pt,                  
  showspaces=false,                
  showstringspaces=false,
  showtabs=false,                  
  tabsize=2,
  float=!htpb,
  xleftmargin=10pt,
  aboveskip=0pt,               
  belowskip=0pt,
}

\setminted{fontsize=\small,
    xleftmargin=6mm,
   linenos,
   frame=lines,
   breaklines}
\begin{document}


\title{An Investigation on Numerical Bugs in GPU Programs Towards Automated Bug Detection}

\author{Ravishka Rathnasuriya}
\orcid{0009-0005-6129-2865}
\affiliation{%
  \institution{University of Texas at Dallas}
  \city{Richardson}
  \country{USA}
}
\email{rsr200002@utdallas.edu}

\author{Nidhi Majoju}
\orcid{0009-0002-2241-0126}
\affiliation{%
  \institution{University of Texas at Dallas}
  \city{Richardson}
  \country{USA}
}
\email{nidhi.majoju@utdallas.edu}

\author{Zihe Song}
\orcid{0009-0009-6651-1560}
\affiliation{%
  \institution{University of Texas at Dallas}
  \city{Richardson}
  \country{USA}
}
\email{zihe.song@utdallas.edu}

\author{Wei Yang}
\orcid{0000-0002-5338-7347}
\affiliation{%
  \institution{University of Texas at Dallas}
  \city{Richardson}
  \country{USA}
}
\email{wei.yang@utdallas.edu}

\input{tex/abstract}

\maketitle
\input{tex/introduction}
\input{tex/background}

\input{tex/study}
\input{tex/ICSE_Debugging}
\input{tex/bugDetection}

\input{tex/bugfixing}

\input{tex/threat}

\input{tex/related}

\input{tex/conclusion}

\clearpage

\bibliographystyle{ACM-Reference-Format}
\bibliography{bibliography}

\end{document}

%% file: tex/abstract.tex
\begin{abstract}

General-purpose graphics processing unit (GPU) computing has emerged as a leading parallel computing paradigm, offering significant performance gains in various domains such as scientific computing and deep learning. However, GPU programs are susceptible to numerical bugs, which can lead to incorrect results or crashes. These bugs are difficult to detect, debug, and fix due to their dependence on specific input values or types and the absence of reliable error-checking mechanisms and oracles. Additionally, the unique programming conventions of GPUs complicate identifying the root causes of bugs, while fixing them requires domain-specific knowledge of GPU computing and numerical libraries. Therefore, understanding the characteristics of GPU numerical bugs~(GPU-NBs) is crucial for developing effective solutions.


In this paper, we conduct a comprehensive study of GPU-NBs by analyzing 397 real-world bug samples from GitHub. We identify common \textit{root causes}, \textit{symptoms}, \textit{input patterns}, \textit{test oracles} that trigger these bugs and the strategies used to \textit{fix} them. \textcolor{black}{We also present GPU-NBDetect, a preliminary tool designed to detect numerical bugs across six distinct bug categories. GPU-NBDetect detected a total of 226 bugs across 186 mathematical
functions in four libraries, with 60 of the bugs confirmed by developers.} Our findings lay the groundwork for developing detection and prevention techniques for GPU-NBs and offer insights for building more effective debugging and auto-repair tool.

\end{abstract}

\ccsdesc[500]{Software and its engineering~Software testing and debugging}
\ccsdesc[500]{Software and its engineering~Empirical software validation}

\keywords{CUDA programming, fault localization, numerical bugs, floating-point errors, bug detection}

%% file: tex/introduction.tex
\section{Introduction}

Numerical applications heavily depend on floating-point arithmetic, which approximates real numbers across varying precision levels, including half, single, and double precision ~\cite{wikipedia_2022}. GPUs are engineered to efficiently execute floating-point operations, offering substantial improvements in computational performance. Notably, NVIDIA GPUs~\cite{nvidiagpus} have progressively advanced with each hardware generation. Early NVIDIA architectures supported only single precision, but modern architectures now fully implement IEEE standards for half, single, and double precision, along with Fused-Multiply-Add (FMA) operations introduced in the Fermi~\cite{fermi} and Kepler~\cite{kepler} architectures. These advancements have introduced distinct characteristics in GPU computations compared to CPU computations, including unique rounding mechanisms, specialized mathematical libraries, and differing numerical capabilities (e.g., compute capability versions 2.0, 3.0, 6.x). 

There are numerous studies on non-numerical bugs in GPU programs and numerical bugs in CPU programs, but none specifically address GPU numerical bugs (GPU-NBs). Existing research in the GPU domain has primarily focused on detecting synchronization bugs~\cite{wu2020simulee, barracuda, gklee, gpuverify, boyer2008automated}, memory bugs~\cite{boyer2008automated, gklee, islam2018bugaroo, sorensen2016exposing}, and compiler bugs~\cite{jiang2020cudasmith}. However, GPU-NBs represent a distinct category arising from the finite approximation of real numbers in computing systems. These bugs can occur due to violations of mathematical properties, numerical operator implementation errors, insufficient data type support, or floating-point representation errors such as round-off and truncation errors. While both numerical and non-numerical bugs can produce similar symptoms, their causes differ. For instance, synchronization bugs result from misuse of barrier APIs~\cite{wu2020simulee} causing thread misbehavior, whereas numerical bugs stem from floating-point representation errors.

Despite the substantial body of research on CPU numerical bugs, a dedicated study of GPU-NBs is necessary due to critical distinctions in their manifestation, detection, debugging and repair~\cite{zou2019detecting,wang2016detecting,zou2015genetic,tang2010perturbing,barr2013automatic,Franco,fu2019effective,DLbug,yi2019efficient,DLGPU2,DLGPU3}. First, GPU and CPU architectures differ fundamentally~\cite{cudahandbook}, affecting how numerical bugs emerge and are triggered. For instance, unlike CPUs, GPUs lack automated fault tolerance mechanisms such as exception flags and  exception status registers~\cite{binfpe, cudaprogrammingguide}, which forces developers to use manual techniques (e.g., insert conditional statements) to identify and diagnose errors. Second, CPUs often perform floating-point calculations using an 80-bit ``extended'' precision mode, whereas GPUs typically use 32-bit and 64-bit for single and double precision operations, respectively~\cite{person_2012}. This disparity in precision impacts how rounding-off errors develop, making testing approaches for CPU numerical  bugs ineffective for GPUs, which round more frequently ~\cite{admin_2013} and often lack the hardware-level rounding controls~\cite{whitehead2011precision} available in CPUs. Third, CPUs typically process operations sequentially, facilitating precise tracking of error accumulation, whereas GPUs run thousands of threads concurrently, leading to complex error patterns that require different debugging and detection approaches. Due to these differences, CPU numerical bug detection strategies~\cite{zou2019detecting,wang2016detecting,zou2015genetic,tang2010perturbing,barr2013automatic,Franco,fu2019effective,yi2019efficient,DLGPU2,DLGPU3,solt2024cascade,ma2024symbolic} cannot be directly applied or work effectively in GPU-NB detection.
Additionally, CPU strategies for numerical bug fixes often involve simple precision adjustments, but GPU fixes demand custom kernel strategies and precision tuning to maintain performance under parallel execution.
Lastly, the proprietary nature of accelerated libraries used in GPU programs~\cite{cufft, cublas, Math, curand,cusolver, cusparse, amgx, cutensor} restricts access to source code, preventing the use of tools like Clang/LLVM for instrumentation~\cite{binfpe}. These factors underscore the need for a focused study on GPU-NBs, examining their symptoms, root causes, and fixing strategies. Such research can improve developers' understanding, and guide future efforts in automated bug detection for GPUs.


\textcolor{black}{Numerical bugs in GPU programs and deep learning (DL) applications originate from different sources. DL numerical issues arise from flaws in the design and logic of algorithms, such as unstable optimization methods or improper weight initialization strategies, affecting prediction accuracy at the application level~\cite{DLGPU1,DLGPU2,DLGPU3,DLGPU4,DLGPU6,DLGPU7}.  In contrast, GPU-NBs stem from how data types and numerical operations are handled during execution, leading to errors at the hardware level. Rather than analyzing entire applications, our focus is on the numerical correctness of core GPU computations and library implementations.} 

\textcolor{black}{DL-based detection approaches rely on whole-application testing methods, such as metamorphic and differential testing~\cite{DLGPU1,DLGPU3}, which perturb models and compare outputs to detect inconsistencies. However, these methods fail to isolate hardware-level numerical errors, as they would require exhaustive exploration of vast model and parameter spaces for whole-application testing. The absence of hardware-level detection techniques leaves a critical gap, as GPU-NBs introduce hidden computation errors without immediate failure signals, making them undetectable through DL-based detection methods.}

The purpose of our study is to investigate GPU-NBs, focusing on common issues and solutions related to floating-point operations. We analyzed bug reports from both NVIDIA's open-source and closed-source numerical libraries, manually examining 906 GPU bug samples, of which 397 were identified as numerical issues. Based on these findings, we categorized numerical bugs into four main types: precision bugs, special-value bugs, device-specific bugs, and data type and operator bugs. We discuss the characteristics of each category, including their \textit{symptoms, test input patterns, oracles} that trigger the bugs, and the strategies used to \textit{fix} these numerical issues in GPU programs.


Our study offers findings and insights that pave the way for future researchers to develop a systematic framework for addressing GPU-NBs. The framework begins with bug detection (Section~\ref{sec: manifestation}), encompassing test input generation and the selection of test oracles. Based on symptoms identified during detection, the framework advances to debugging (Section~\ref{sec:root-cause}) by mapping symptoms to potential root causes (Fig.~\ref{fig:Overviewfinal}). Once the root causes are identified, the framework applies corresponding fixing strategies (Fig.~\ref{fig:Overviewfinal}) to repair the program (Section~\ref{sec: bugfixing}). We propose a preliminary tool, GPU-NBDetect, for detecting GPU-NBs (Section~\ref{subsec: tool}). GPU-NBDetect employs seven mutation rules to generate floating-point inputs and detects numerical bugs across six categories. GPU-NBDetect identified 226 bugs across 186 mathematical functions in four libraries, with 60 confirmed by developers, outperforming four existing bug detectors and two input generation methods.

Our main contributions of this study are:
\begin{itemize}[leftmargin=*]
\item Summarizing patterns of the symptoms, the root causes, and the fixing strategies of numerical bugs in GPU programs. 
\item Providing guidance for creating reliable and stable GPU program test inputs and test oracles for detecting numerical bugs.
\item Outlining common fix strategies for numerical issues, helping developers reproduce and resolve bugs efficiently, thus conserving development resources.  
\item Offering insights into the automated detection, debugging, and fixing of GPU-NBs.
\item Introducing GPU-NBDetect, a preliminary tool featuring input generation and test oracle selection to detect GPU-NBs.
\end{itemize}

\begin{figure}[htbp]
\centering
\includegraphics[width=\columnwidth]{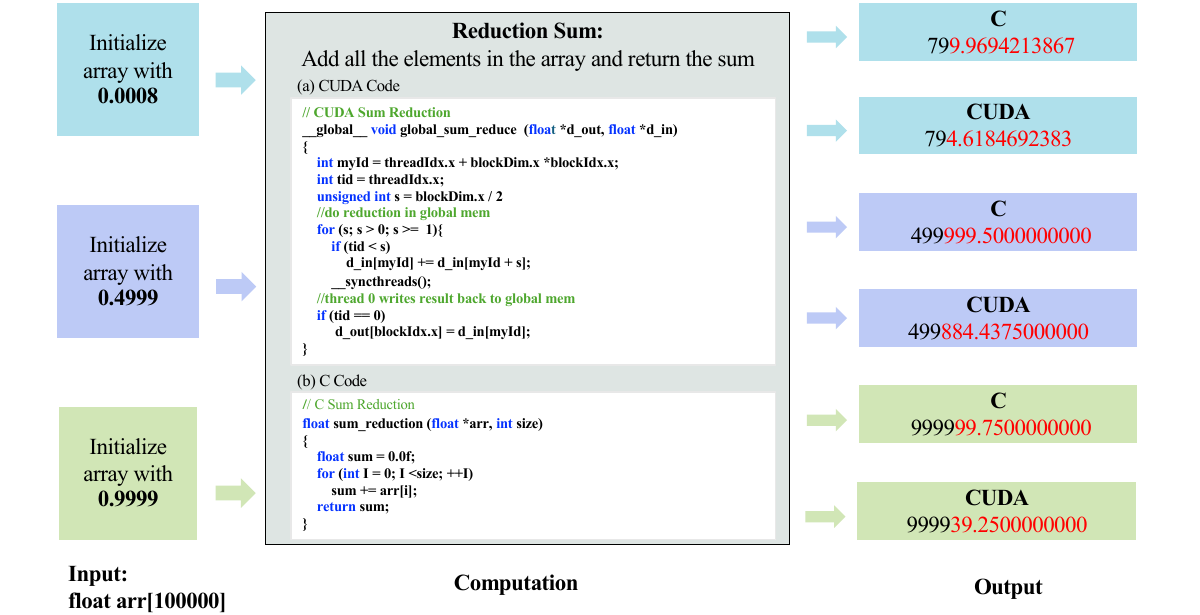}
\caption{Motivating Example: CUDA Produces Inconsistent Result for Reduction Sum.}
\label{motivation}
\end{figure}

\input{tex/tables/findings-implications}

%% file: tex/tables/findings-implications.tex
\begin{table}[hbpt!]
\caption{Summary of Findings (F) and Implications (I)}
\label{tab:Findings-Implications}
\resizebox{\linewidth}{!}{%
\begin{tabular}{|l|p{7cm}|p{12cm}|}
\hline
{\bf } & {\bf Findings about numerical bug causes}  & {\bf Implications on unique characteristics of GPU-NBs} \\ \hline

1 & Among the 397 observed GPU program tests, 97 bugs were identified as precision bugs, including overflow/underflow, rounding-off, and casting errors. (Table~\ref{table:root_Cause}) & Precision bugs in GPUs occur because, unlike CPUs with extended precision (80-bit mode) and automatic error detection, GPUs lack mechanisms to control precision, causing errors to propagate unnoticed.
\\\hline


2 & One-third of precision issues are due to rounding errors (Table~\ref{table:root_Cause}).   & Rounding errors arise because GPUs lack programmable rounding modes and extended precision support. Unlike CPUs, which allow flexible precision adjustments, GPUs use fixed rounding mechanisms, leading to cumulative precision loss, particularly in parallel computations.
 \\ \hline
 3 & One-third of precision issues are due to casting errors (Table~\ref{table:root_Cause}). Studies on CPU programs have not identified casting errors.  &  Casting errors occur because, unlike CPUs, GPUs substitute default values for unsupported data types rather than raising exceptions, allowing errors to go undetected.
 \\\hline
 
 

 4 & Among the 397 observed GPU program tests, 239 (59.6\%) are caused by data type and operator issues unique to GPU, particularly with CUDA numerical operations (Table~\ref{table:root_Cause}). & GPU software bugs stem from unsupported data types and unimplemented operations in CUDA, as limited transparency and testing in GPU libraries leave compatibility issues undetected, unlike fully verified CPU libraries.
 \\\hline


5 & Special value bugs account for 12.1\% of the observed GPU program tests (Table~\ref{table:root_Cause}). & GPU bugs arise from a lack of automatic NaN and INF detection, causing silent accumulation of rounding errors. Unlike CPUs with built-in exception alerts, GPUs rely on manual checks to handle these inconsistencies.
\\\hline



6 & Environmental issues account for 3.7\% of numerical problems in GPU programs and are unique to GPU (Table~\ref{table:root_Cause}). & GPU bugs stem from inconsistent precision support across device generations. Designed for speed over accuracy, GPUs lack uniform numerical support, forcing developers to manually manage precision inconsistencies.
\\\hline 


{\bf } & {\bf Findings about numerical bug manifestation}  & {\bf Implications on bug detection} \\ \hline 
7 & Developers use floating-point numbers, integers, out-of-range values, and NaNs as inputs to GPU programs (Section \ref{subsec: input}). & Automated, feedback-driven fuzzing improves GPU bug detection by dynamically generating test inputs that expose numerical bugs, addressing GPU's sensitivity to data types and rounding behavior.
\\\hline 

8 & All GPU programs can be tested against their corresponding CPU versions  (Fig.~\ref{fig:Overview}). & Automated cross-platform testing enhances GPU bug detection by using CPU benchmarks to reveal GPU-specific discrepancies, leveraging the CPU's robust precision control.
\\\hline

9 & Of the 397 tests, 179 (45.1\%) have the capability to log program results  (Tables~\ref{table:symptoms} and~\ref{table:ManifestationStrategies}). & Automated logging aids GPU bug detection by capturing results and comparing them with CPU outputs, effectively identifying silent rounding and casting issues.
\\\hline 

10 & Of the 397 tests, 37 (9.3\%) can profile program performance in terms of execution (Tables~\ref{table:symptoms} and~\ref{table:ManifestationStrategies}). & 
Automated performance profiling can enhance GPU bug detection by flagging bottlenecks linked to precision-speed trade-offs, which are less pronounced in CPU.
\\\hline

{\bf } & {\bf Findings about numerical bug symptoms}  & {\bf Implications on debugging} \\ \hline 
11 & The observed GPU program tests exhibit three primary symptoms: error messages (45.6\%), wrong results (45.1\%), and performance bottlenecks (9.3\%) (Fig.~\ref{fig:Overviewfinal} and Table~\ref{table:symptoms}). & Debugging GPU-NBs reveals symptoms closely tied to root causes (Fig.~\ref{fig:Overviewfinal}). Unlike CPUs, GPUs experience more severe performance bottlenecks from numerical bugs, often leading to significant performance losses (Table~\ref{table:ManifestationStrategies}).
\\\hline

12 & Error messages are prominent in GPU programs, occurring in 181 out of 397 cases (Table~\ref{table:symptoms}). & Unique error messages in GPUs such as overflow/underflow, invalid casting, unsupported data types, operators, and undefined computations can directly reveal root causes (Fig.~\ref{fig:Overviewfinal}). Unlike CPUs with fault-tolerant mechanisms like exception detection, GPUs require developers to manually insert such checks for effective debugging.
\\\hline

13 & Producing wrong results helps identify numerical bugs in GPU programs. Our study found that 179 out of 397 issues (45.1\%) were confirmed by comparing program results (Table~\ref{table:symptoms}). & Debugging wrong results in GPUs involves comparing outputs across scenarios, such as varying precision levels, casting operations, data types, and checking for NaN values (Fig.~\ref{fig:Overviewfinal}).
 \\\hline 

14 & Performance bottlenecks help detect numerical bugs even when programs produce correct results, occurring in 9.3\% of cases (Table~\ref{table:symptoms}). & Debugging performance bottlenecks in GPUs involves comparing cast operations, data types, and CPU execution times, with visual profilers aiding analysis of time and memory usage (Fig.~\ref{fig:Overviewfinal}).
\\\hline

{\bf } & {\bf Findings about numerical bug fixing}  & {\bf Implications on program repair} \\ \hline

15 & Various precision issues are addressed through different methods: 27.8\% with checkers, 39.3\% by changing precision, and 32.9\% using upcasting/downcasting strategies  (Table~\ref{table:fixing strategies} and Fig.~\ref{fig:Overviewfinal}). & Automated repair approaches for precision issues could save time by adjusting numerical precision without relying on manual, trial-and-error methods, which are typically time-consuming for users.
 \\\hline

16 & Four out of 57 (7.1\%) operator issues in third-party libraries are resolved using custom kernel strategies, which are unique to GPU programs (Table~\ref{table:fixing strategies} and Fig.~\ref{fig:Overviewfinal}). & CuPy's custom kernel strategy assists users in defining numerical operators within GPU programs.  
\\\hline

17 & Data type issues in GPU programs can only be resolved by adding data type support or using a compatible data type (Table~\ref{table:fixing strategies} and Fig.~\ref{fig:Overviewfinal}). & Automated repair could address data type issues by adding support for required types or selecting compatible data types, reducing the need for manual adjustments. 

 \\ \bottomrule
\end{tabular}
}
\end{table}

%% file: tex/background.tex
\section{Background and Motivation}



Floating-point numbers approximate real values based on the IEEE-754 standard~\cite{wikipedia_2022}. GPUs support multiple formats: half, single, double, tensor and brain floating point~\cite{gopalakrishnan2021guarding}, making format selection critical for ensuring computational accuracy in applications ranging from scientific computing to DL.





While GPUs excel at parallel computation, they are prone to numerical inaccuracies, particularly in floating-point arithmetic. Fig.~\ref{motivation} highlights discrepancies in reduction sum operation results between CPU and GPU computations for values like 0.0008 and 0.4999. For example, the CPU computes 799.969, while CUDA yields 794.618; similarly, 499999.500 on the CPU becomes 499884.438 on the GPU. These deviations stem from cumulative rounding errors~\cite{admin_2013,cudahandbook,gopalakrishnan2021guarding} and non-associative nature of floating-point operations, reflecting the precision limitations of GPU computations. 

These examples highlight a broader issue: \textit{as computation complexity increases, GPU-NBs can cause substantial deviations from expected results}, which is critical for accuracy-dependent applications. Despite the manual developer efforts to adjust precision or error-checking mechanisms, the root causes of GPU-specific inaccuracies remain insufficiently explored. This study offers a comprehensive analysis of GPU-NBs, examining real-world cases to identify root causes, symptoms, input patterns, bug-triggering oracles, and effective mitigation strategies.

%% file: tex/study.tex
\section{Methodology}
\label{method}
In this section, we describe our data collection process (Section~\ref{data_collection}) and discuss the methodology for filtering and labeling GPU-NBs (Sections~\ref{selection} and~\ref{labelling}).

\subsection{Data Collection}
\label{data_collection}


Building a comprehensive dataset of GPU-NBs requires systematic identification of relevant issues across libraries and projects. Since GPU libraries lack standardized popularity metrics like power scoring~\cite{Hales_2018}, we used GitHub as the primary source, collecting bug reports from January 2020 to June 2024 through a combination of automated queries and manual review to ensure relevance.



Our methodology distinguishes between closed-source and open-source GPU libraries. For closed-source libraries, such as cuFFT~\cite{cufft}, cuBLAS~\cite{cublas},  Math~\cite{Math}, cuRAND~\cite{curand}, cuSOLVER~\cite{cusolver}, cuSPARSE~\cite{cusparse}, and cuTENSOR~\cite{cutensor}, we collected bug samples from dependent projects due to limited direct access. In contrast, open-source libraries like CuPy~\cite{cupy}, Numba~\cite{numba}, MatX~\cite{matx}, and Ginkgo~\cite{ginkgo} allowed direct analysis of the libraries and their associated projects. This approach enabled the construction of a diverse GPU-NB dataset spanning domains from scientific computing to DL.

\subsection{Issue Selection}
\label{selection}

Identifying GPU-NBs across GitHub is time and labor-intensive, therefore, we adopted a systematic keyword-based filtering approach tailored to both closed- and open-source libraries. This section outlines the criteria and methods used to extract relevant bugs and eliminate invalid reports.

\textbf{Selection GPU-NBs related to closed-source libraries.} Using the GitHub search API~\cite{gharchive}, we initially collected 118,438 reports with broad keywords such as ``GPU'' and ``GPU programming''. To focus on numerical bugs, we applied a refined set of keywords based on methodologies from previous studies on CPU numerical bugs~\cite{Franco}. These keywords included terms such as ``nan'', ``overflow'', ``underflow'', ``infinity'', ``precision'', ``unstable'', ``roundoff'', ``truncation'', ``cancellation'', ``accuracy'', ``numeric'', ``floating'', ``fp'', ``cufft'', ``cublas'', ``curand'', ``cuda math library'', ``cusolver'', and ``cusparse''.  This filtering process narrowed our sample to 10,380 issues.

\textit{Removing Invalid Issues}: To ensure relevance and reduce noise, we applied further criteria: (1) confirmed bugs (closed status), (2) consists of a documented fix, and (3) classified as a numerical bug (excluding memory and synchronization issues). These steps narrowed the sample size to 624.

\textbf{Selection GPU-NBs related to open-source libraries.} For open-source libraries, we expanded the search to include both the libraries and their dependent projects. Using terms like ``CUDA'' and ``numerical'', we initially identified 76 repositories. We then combined ``CUDA'' with key numerical libraries from Franco et al. ~\cite{Franco}, yielding additional repositories for ``NumPy'', ``SciPy'', ``LAPACK'', ``GSL'', and ``Elemental''. After removing duplicates and irrelevant projects, we finalized a set of open-source libraries and projects, including CuPy~\cite{cupy}, Numba~\cite{numba}, MatX~\cite{matx}, and Ginkgo~\cite{ginkgo}.
\begin{figure}[htbp]
\centering
\includegraphics[width=0.9\columnwidth]{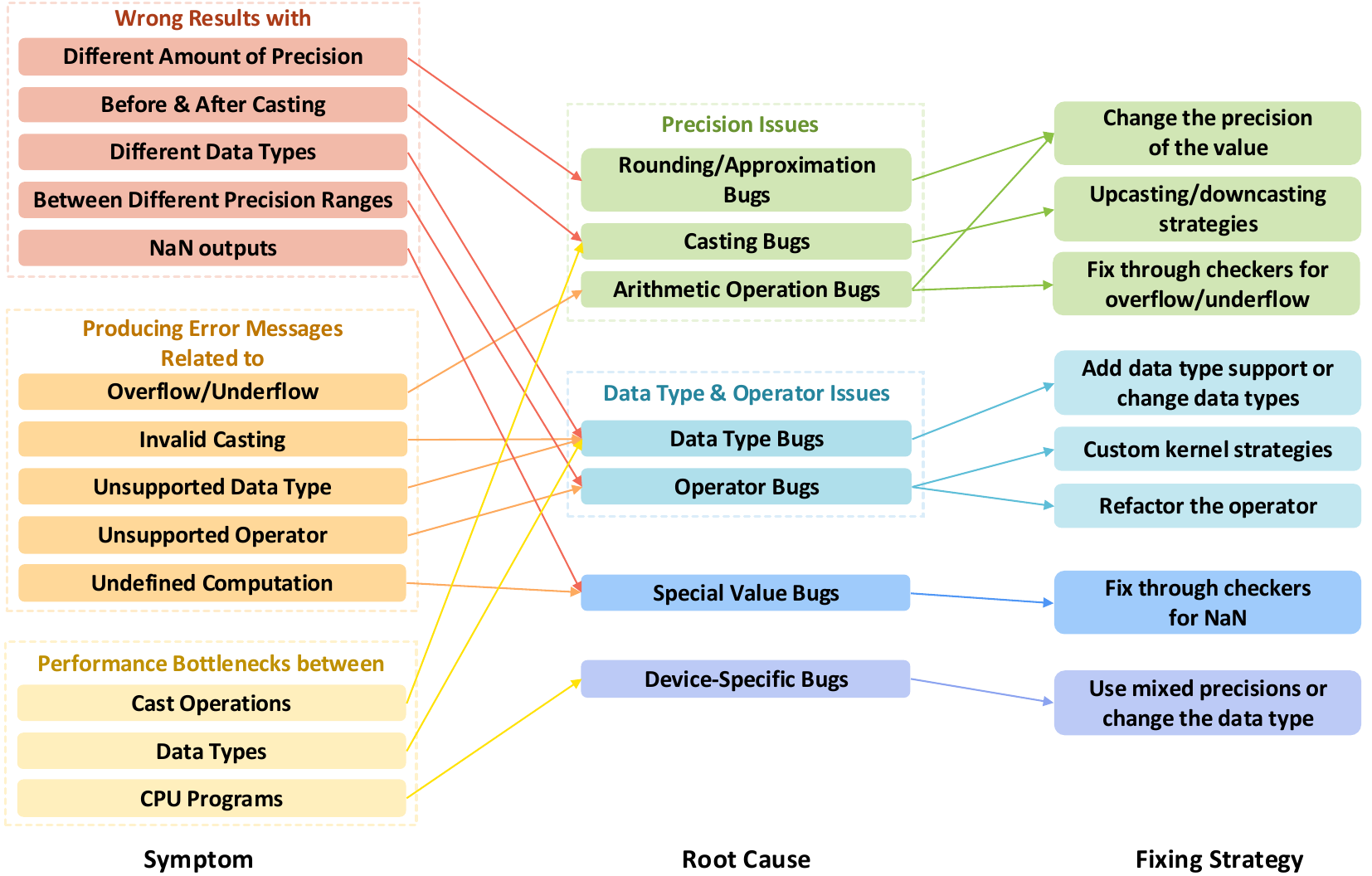}
\caption{Relationship among symptoms, root causes, and fixing strategy}
\label{fig:Overviewfinal}
\end{figure}

\textit{Removing invalid issues}:  We refined the dataset by retaining only closed bug reports, as open reports may reflect unresolved or unverified bugs. This final filtering yielded 282 validated GPU-NB cases.

\subsection{GPU-NB Labeling}
\label{labelling}
After merging both datasets, removing duplicates, and excluding misclassified bug reports, we obtained 397 unique GPU-NB cases (229 from closed-source and 168 from open-source libraries). These were categorized by root cause using a three-step process.


First, two authors independently reviewed each bug report by analyzing descriptions, comments, and associated commits. While we initially applied Franco et al.'s CPU-focused framework~\cite{Franco}, it lacked specificity for GPU-related issues. Notably, the CPU categories for \textit{accuracy} and \textit{convergence} overlapped, as both addressed round-off and truncation errors. To resolve this, we merged them into a broader precision bugs category, which we expanded to include \textit{arithmetic}, \textit{approximation}, and \textit{casting} bugs unique to GPUs. Additionally, we introduced two new categories: \textit{data type and operator} bugs and \textit{device-specific} bugs to capture GPU-architecture-specific issues. This refinement resulted in seven distinct categories, with only \textit{special value} bugs overlapping with CPU studies.

Second, we ensured consistency by comparing initial categorizations and merging similar categories as needed. A joint review was conducted to resolve 25 inconsistently labeled reports.

Finally, we mapped specific symptoms to each root cause, enabling a fine-grained categorization of GPU-NB manifestations. Unlike the broader symptom types used in CPU studies, our detailed classification captures GPU-specific behaviors and reveals strong one-to-one correspondences between symptoms, root causes, and fixes. We further analyzed common repair strategies, noting a reliance on manual fixes in GPU development, in contrast to the automation observed in CPU-focused work. The complete mapping is illustrated in Fig.~\ref{fig:Overviewfinal}, and a summary of the analyzed GitHub reports is available on our project website~\cite{GPUstudy}.

%% file: tex/ICSE_Debugging.tex
\section{Causes of Numerical Bugs}
\label{sec:root-cause}



We identified 397 numerical bugs, which we categorized into four primary groups:  ~\textit{precision, data type and operator, special value,} and ~\textit{device-specific} bugs. Table~\ref{table:root_Cause} presents the distribution of each bug type across these categories.




\input{tex/tables/root_cause_category_stats}

\subsection{Precision Bugs}

Precision bugs in GPU programs arise from limitations in representing floating-point numbers accurately, leading to computations with insufficient or imprecise numerical values. Approximately 24.4\% of the identified numerical bugs fall into this precision category (I-1 in Table ~\ref{tab:Findings-Implications}). These bugs fall into one of three subcategories:




\emph{(1) Arithmetic Operation Bugs: }
Arithmetic operation bugs arise when computations exceed or fall below the supported numerical range, resulting in overflow or underflow. Unlike CPUs, which use exception status registers (e.g., \textit{fenv.h} in C) to detect such conditions, GPUs lack built-in mechanisms for numerical exception handling. As a result, computations may silently yield values like +inf, -inf, or 0.0~\cite{binfpe} unless developers manually implement detection logic.

These bugs are particularly prominent in GPU programs due to the frequent use of mixed-precision computation~\cite{tensorflowbackground2, tensorflowbackground1} and integer quantization~\cite{tensorflowbackground3}. For example, values stored in FP32 may be processed in FP16, increasing the risk of range violations. Additionally, implicit conversions from integers to floating-point types can introduce further precision loss and memory overhead. As shown in Fig.~\ref{fig:Overviewfinal}, arithmetic operation bugs in GPU code often lead to specific error messages, such as developer-inserted overflow or underflow flags.

\begin{figure}[htbp]
    \centering
    \begin{subfigure}{0.4\linewidth}
        \captionsetup[lstlisting]{singlelinecheck=false, justification=raggedright, labelfont=bf}
\begin{lstlisting}[language=Python,basicstyle=\tiny,breakatwhitespace=true,postbreak=\mbox{\textcolor{red}{$\hookrightarrow$}\space}]
# Scipy's convolve2d
import numpy as np
from scipy.signal import convolve2d
a = np.array([[257]], dtype="uint64")
b = np.array([[1]], dtype="uint8")
result = convolve2d(a, b, mode="same")
print(result, result.dtype)
# [[257]] uint64 : expected result
\end{lstlisting}
        \caption{SciPy Version with Correct Behavior }
        \label{fig: AccuracyOverflow-a}
    \end{subfigure}
    \hfill 
    \begin{subfigure}{0.58\linewidth}
\begin{lstlisting}[language=Python,basicstyle=\tiny,breakatwhitespace=true,postbreak=\mbox{\textcolor{red}{$\hookrightarrow$}\space}]
# Cupy's convolve2d
import cupy as cp
from cupyx.scipy.signal import convolve2d
a = cp.asarray(a)
b = cp.asarray(b)
result = convolve2d(a, b, mode="same")
print(result, result.dtype)
# [[1]] uint64 : correct dtype, but uint8 integer overflow
# Reversed convolution
result = convolve2d(b, a, mode="same")
print(result, result.dtype)
# [[1]] uint8: wrong dtype and uint8 integer overflow
\end{lstlisting}
        \caption{CuPy Produces an Integer Overflow}
        \label{fig: AccuracyOverflow-b}
    \end{subfigure}
    \caption{An Instance of Precision Bug due to an Integer Overflow in {\tt CuPy~\cite{Accuracy2}}. (a) SciPy Version with Correct Behavior. (b) CuPy Produces an Integer Overflow.}
    \label{fig: AccuracyOverflow-eg}
\end{figure}

An example from the CuPy project~\cite{Accuracy2} illustrates this problem in Fig.~\ref{fig: AccuracyOverflow-eg}. During the convolve2d operation, an integer overflow occurs because the operation is executed with incompatible data types. When applied to two integer arrays, $a$ and $b$, with different precisions (uint64 and uint8, as shown in lines 4 and 5 in Fig.~\ref{fig: AccuracyOverflow-a}), the SciPy library correctly handles the operation in the higher precision data type (uint64), avoiding overflow (line 6). In contrast, CuPy defaults to the lower precision type, uint8 (line 15 in Fig.~\ref{fig: AccuracyOverflow-b}), resulting in integer overflow since uint64 values cannot be accurately represented in uint8.

\emph{(2) Rounding/Approximation Bugs: }Bugs in this category occur when real numbers are approximated within finite precision limits, causing deviations from expected values. While both CPU and GPU programs are affected, the impact is more pronounced on GPUs due to their limited precision and lack of extended precision support. Precision loss is further exacerbated by operation ordering effects such as subtractive cancellation~\cite{substractive}. 





CPUs often mitigate such errors using 80-bit extended precision, whereas GPUs are restricted to 32-bit or 64-bit precision and lack configurable rounding modes~\cite{cudahandbook, gopalakrishnan2021guarding} (I-2 in Table~\ref{tab:Findings-Implications}). Instead, developers must rely on compiler intrinsics to control rounding behavior. As shown in Fig.~\ref{fig:Overviewfinal}, rounding bugs in GPUs can lead to significant discrepancies between actual and expected outputs.

\begin{figure}[htbp]
    \centering
    \begin{minipage}{0.5\linewidth}
        \captionsetup[lstlisting]{singlelinecheck=false, justification=raggedright, labelfont=bf}
\begin{lstlisting}[language=Python,basicstyle=\tiny,breakatwhitespace=true,postbreak=\mbox{\textcolor{red}\space}]
arr = cp.arange(12.).reshape((4, 3))
points = cp.array([[0.5, 2], [0.5, 1]]) 
# Scipy defaults to order=3
# But cupy gives only zeros
print(f"order={None}->", cupy_map_coordinates(arr, scipy_map_coordinates(arr.get(), points.get()))
for order in [0, 1]:
    print(f"order={order} ->", cupy_map_coordinates(arr, points,order=order), scipy_map_coordinates(arr.get(), points.get(), order=order))
\end{lstlisting}
        \caption{An Instance of Precision Bug due to Rounding Bug in in {\tt CuPy~\cite{Accuracy1}}}
        \label{fig: Accuracy-eg}
    \end{minipage}
    \hfill
    \begin{minipage}{0.46\linewidth}
        \captionsetup[lstlisting]{singlelinecheck=false, justification=raggedright, labelfont=bf}
        \centering
\begin{lstlisting}[language=Python,basicstyle=\tiny,breakatwhitespace=true,postbreak=\mbox{\textcolor{red}\space}] %{$\hookrightarrow$}
import cudf
from cudf import _lib as libcudf
from cudf.core.dtypes import Decimal64Dtype
import numpy as np
ser = cudf.Series([-72462.12]) .astype(Decimal64Dtype(10,5))
libcudf.unary.cast(ser._column, np.dtype("int32"))
# Result: 13437
ser = cudf.Series([-48938], dtype="int32")
libcudf.unary.cast(ser._column, Decimal64Dtype(10,5))
# Result: -5988.32704
\end{lstlisting}
        \caption{An Instance of Precision Bug due to a Casting Error in {\tt cuDF~\cite{casting}}}
        \label{fig: casting-eg}
    \end{minipage}
\end{figure}


An example of this can be seen in the CuPy project ~\cite{Accuracy1} in Fig.~\ref{fig: Accuracy-eg}, where the {\tt map\_coordinate} operation with order=none and order=0 produces incorrect outputs compared to SciPy. In this case, the input array {\tt point} in line 2 contains a value of 0.5, leading to a rounding error in CuPy's precision handling. CuPy uses the {\tt rint()} function, which rounds to the nearest integer, whereas SciPy employs {\tt floor(double)}, which rounds down to the nearest double value. This difference in rounding methods leads to output inconsistencies between the two libraries.

\emph{(3) Casting Bugs:} 
Casting bugs occur when values are converted between data types, resulting in numerical inconsistencies. While relatively rare in CPU numerical libraries ~\cite{Franco}, casting issues are more prevalent in GPU programs, where such operations often produce inconsistent or erroneous outcomes (I-3 in Table~\ref{tab:Findings-Implications}).





Casting between high and low precision data types introduces risks of precision loss, which should ideally remain within acceptable bounds (e.g., within 0.1). Unlike CPUs that leverage register flags to detect casting errors, GPUs lack exception-handling mechanisms. As a result, when a value is cast to an unsupported type, GPUs often substitute default values, allowing errors to propagate silently and undetected.

As shown in Fig.~\ref{fig:Overviewfinal}, casting issues in GPUs can be detected by (1) examining discrepancies between numerical values before and after casting, and (2) identifying performance bottlenecks when converting between precision types.

An example of a casting issue is shown in the cuDF project~\cite{casting}, illustrated in Fig.~\ref{fig: casting-eg}.
In this program, the $unary.cast$ operation between Decimal64 and int32 yields erroneous results (lines 6 and 9). The $unary.cast$ operator lacks an exception detection mechanism, failing to verify whether the source value can be accurately represented in the target data type, int32\_t. Consequently, casting -72462.12 from Decimal64 to int32 produces an inaccurate default value of 13437, while casting -48938 from int32 to Decimal64 results in -5988.32704.

\begin{figure}[!htbp]
    \centering
    \begin{subfigure}[t]{0.49\columnwidth}
       \includegraphics[width=\linewidth]{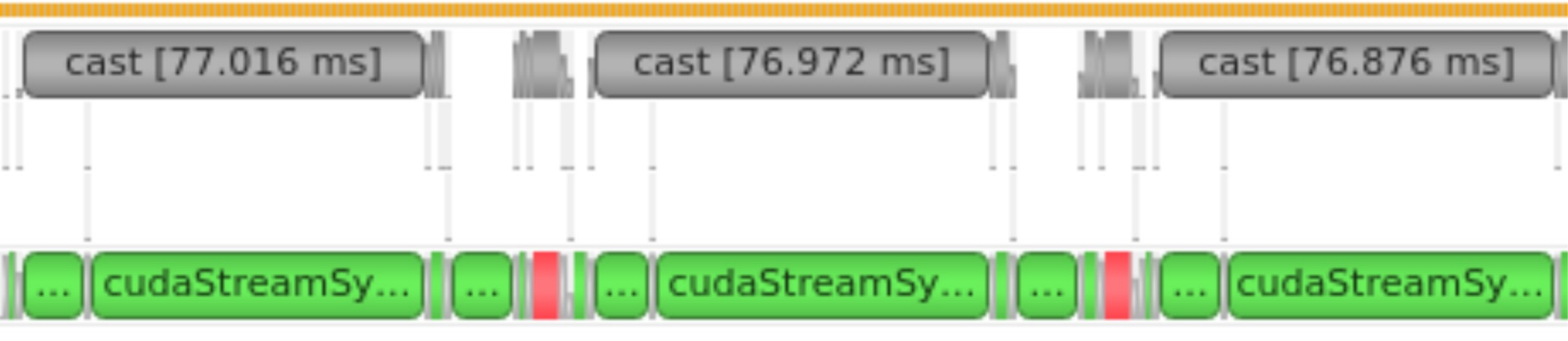}
       \caption{Input type of Decimal64 is casted to Decimal128 as output shows slow performance in {\tt cuDF~\cite{castVisual}}}
       \label{fig:Ng1}
    \end{subfigure}
    \hfill
    \begin{subfigure}[t]{0.49\columnwidth}
       \includegraphics[width=\linewidth]{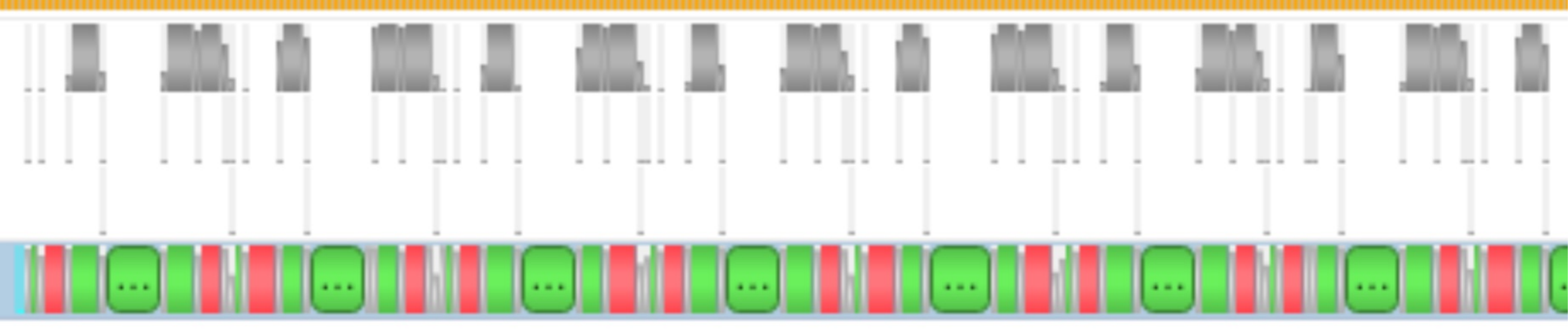}
       \caption{Input type of Decimal128 is output as the same with faster performance in {\tt cuDF~\cite{castVisual}}}
       \label{fig:Ng2}
    \end{subfigure}
    
    \caption{Performance Bottleneck When Casting a Data Type {\tt cuDF~\cite{castVisual}}}
    \label{fig:Casting-Performance}
\end{figure}

Furthermore, a cuDF example~\cite{castVisual}, shown in Fig.~\ref{fig:Ng1} and ~\ref{fig:Ng2} illustrates how data type choice affects casting performance.
Casting from Decimal64 to Decimal128 required 4134 ms, while casting between identical Decimal128 types took only 1598 ms, indicating a 2.5-fold difference in execution time. In these figures, grey sections indicate the casting operation duration, and green sections represent CUDA API activities.

\subsection{Data Type and Operator Bugs}
Data types~\cite{datatypedefinition} specify how values are represented, while operators~\cite{operatordef} define how computations are performed on them. Bugs related to data types and operators account for 59.6\% of numerical issues in GPU programs (I-4 in Table~\ref{tab:Findings-Implications}). These bugs typically arise from insufficient support for certain data types or operations, resulting in incorrect behavior or performance degradation.




Data type bugs arise when unsupported or incorrect types are used in computations, often leading to invalid casting or unsupported data type errors (Fig.~\ref{fig:Overviewfinal}). For example, FP16 operations may perform worse than FP32 under certain conditions, causing unexpected performance bottlenecks. Operator bugs result from missing or incorrectly implemented functions, leading to unsupported operations or erroneous outputs. These two categories are often interrelated, as data type issues can be misinterpreted as operator failures.

While GPU libraries often adopt CPU-style implementations for numerical operators, they differ significantly in documentation quality and transparency. Many are closed-source, preventing developers from verifying supported data types or internal behaviors. Consequently, operators that work correctly on CPUs may fail on GPUs due to inadequate type support. This opacity complicates debugging, as users cannot inspect or modify internal implementations. Unlike CPU libraries, which primarily address typographical issues~\cite{Franco}, GPU-NBs more frequently stem from unsupported data types and operator incompatibilities.

\begin{figure}[htbp]
    \centering
    \begin{minipage}{0.4\linewidth}
        \captionsetup[lstlisting]{singlelinecheck=false, justification=raggedright, labelfont=bf}
        \centering
\begin{lstlisting}[language=Python,basicstyle=\tiny,breakatwhitespace=true,postbreak=\mbox{\textcolor{red}{$\hookrightarrow$}\space}]
a = cp.ones(10, dtype=cp.float16)
b = cudf.Series(a)
\end{lstlisting}
        \caption{An Instance of Data Type Issue in {\tt cuDF~\cite{datatype}}}
        \label{fig: datatype-eg}
        \captionsetup[lstlisting]{singlelinecheck=false, justification=raggedright, labelfont=bf}
        \centering
\begin{lstlisting}[language=Python,basicstyle=\tiny,breakatwhitespace=true,postbreak=\mbox{\textcolor{red}{$\hookrightarrow$}\space}]
torch.autograd.set_detect_anomaly(False)
a = torch.Tensor([1e-4]).cuda().half()
a.requires_grad = True
a.norm(3).backward()
a.grad
\end{lstlisting}
        \caption{Producing NaN with FP16 in PyTorch Library in {\tt PyTorch~\cite{special2}}}
        \label{fig: SpecialValue2-eg}
    \end{minipage}
    \hfill
    \begin{minipage}{0.56\linewidth}
        \captionsetup[lstlisting]{singlelinecheck=false, justification=raggedright, labelfont=bf}
        \centering
\begin{lstlisting}[language=Python,basicstyle=\tiny,breakatwhitespace=true,postbreak=\mbox{\textcolor{red}\space}]
# NumPy Version with Correct Behavior
numpy_arr = np.random.random((20, 20))
numpy_result = scipy.ndimage.filters.generic_filter( numpy_arr, np.mean, 3, mode="reflect", cval=0)

# CuPy Produces an Operator Error
cupy_arr = cupy.random.random((20, 20))
cupy_result = cupyx.scipy.ndimage.filters.generic_filter( cupy_arr, cupy.mean, 3, mode="reflect", cval=0)
\end{lstlisting}
        \caption{An Instance of Operator Issue in {\tt CuPy~\cite{correctness1}}}
        \label{fig: Correctness-eg}
    \end{minipage}
\end{figure}



In the cuDF project~\cite{datatype}, an example of a data type bug occurs when creating a series of float16 CuPy objects, causing a crash with an ``invalid data type'' error in Fig.~\ref{fig: datatype-eg}. This issue arises because the {\tt \_\_cuda\_array\_interface\_\_} expects {\tt np.dtype} objects, but the program incorrectly uses cudf.dtype, leading to a failure.

An operator-related issue is observed in the CuPy project~\cite{correctness1} in Fig.~\ref{fig: Correctness-eg}, where the program fails to pass the {\tt cupy.mean} function into the {\tt generic\_filter()} operation, resulting in a ``bad type error'' message. However, the equivalent np.mean function in NumPy passes without issue (line 3), highlighting a compatibility problem with CuPy's handling of certain operators.

\subsection{Special Value Bugs } 
Special value bugs arise when computations produce or encounter NaN or (+/-)INF values, accounting for 12.1\% of the dataset (I-5 in Table~\ref{tab:Findings-Implications}). These bugs often occur when numerical operations yield undefined results, causing error messages or incorrect outputs due to the presence of NaN or INF values (Fig.~\ref{tab:Findings-Implications}).


A primary cause of special value bugs in GPUs is the lack of built-in exception handling for NaN and INF values. Unlike CPUs, which include exception detection for these cases, GPUs require developers to manually implement checks for special values. Without automated detection, errors involving NaN or INF can propagate undetected, resulting in incorrect computations.




Differences in NaN handling between CUDA and CPU-based libraries have also been observed~\cite{special2}. For example, NVIDIA GPUs use little-endian format exclusively, whereas CPU-based libraries like NumPy support both little- and big-endian formats. CuPy, however, limits support to little-endian, which can lead to incorrect results when processing big-endian data~\cite{CupyNaN}.

An example from the PyTorch project~\cite{special2} illustrates a special value bug (Fig.~\ref{fig: SpecialValue2-eg}). NaN values appear in \verb|a.grad| computations with FP16 values (line 5) when a small value is passed to the \verb|torch.tensor()| function. This results in INF values that propagate and produce NaN outputs.

\subsection{Device-specific Bugs}
Device-specific bugs in GPU programs arise from variations in numerical support across different GPU models. As shown in Table~\ref{table:root_Cause}, 15 out of the 397 identified bugs are attributed to device-specific factors (I-6 in Table~\ref{tab:Findings-Implications}). These bugs often result in performance delays compared to CPU execution time when using GPUs with limited numerical support (Fig.~\ref{fig:Overviewfinal}).




While CPU architectures generally maintain backward compatibility and consistent numerical support across generations, GPU support can vary, leading to device-specific issues ~\cite{tensorflowbackground2}. For example, older GPUs lack support for mixed-precision computing, while newer models, such as the RTX series, V100, and A100, achieve substantial performance improvements through mixed precision.

In the CuPy project~\cite{environment1}, performance discrepancies in double precision computations illustrate this issue. Tests reveal that matrix multiplication with NumPy on a CPU is faster than the CuPy implementation on a GPU. Specifically, the CPU execution time with NumPy was 0.7268 seconds, while the GPU execution time with CuPy was 1.0382 seconds. The test was conducted on a GeForce GTX 760 GPU, which is known for its low double precision performance.

\input{tex/tables/symptoms}
\subsection{Mapping Symptoms to Root Causes}

This section summarizes methods for identifying the root causes of GPU-NBs based on their symptoms. Our study identifies three symptom types resulting from GPU-NBs: incorrect results, error messages, and performance bottlenecks, as depicted in Fig.~\ref{fig:Overviewfinal}. Moreover, we extend the classification with detailed subcategories for each symptom type. Table~\ref{table:symptoms} presents the distribution of these symptoms across identified GPU-NBs.



Unique error messages directly indicate the root cause of a GPU-NB, particularly when developers place manual checkers in GPU code due to the lack of exception detection capabilities. For example, overflow or underflow errors suggest that a computed value has exceeded the supported precision range, indicating an arithmetic operation bug. Errors such as ``invalid casting'' occur when a numerical value is cast to an unsupported data type, while ``unsupported data type'' or ``unsupported operator'' errors signal attempts to use data types or operations not supported by the API. Special value bugs are typically flagged by errors involving undefined values like NaN or INF (F-12 and I-12 in Table~\ref{tab:Findings-Implications}).

Comparing GPU and CPU outputs for the same program provides further insight into GPU-NBs. Manually adapting GPU programs to corresponding CPU implementations (e.g., replacing CuPy with NumPy or SciPy) helps detect discrepancies. Casting bugs are confirmed when results differ before and after casting, while precision discrepancies point to rounding bugs. Varying results across different data types for the same values suggest a data type bugs, and differences across precision ranges for specific operators indicate an operator issue. NaN outputs consistently identify special value bugs (F-13 and I-13 in Table~\ref{tab:Findings-Implications}).

For programs yielding correct results, performance measurement can also reveal numerical bugs. Performance bottlenecks often indicate underlying issues; for instance, casting operations can introduce inefficiencies, with lower precision types like FP16 sometimes performing worse than FP32. Programs that show poorer performance on GPUs than CPUs may signal device-specific bugs (F-14 and I-14 in Table~\ref{tab:Findings-Implications}), as detailed in Section~\ref{sec:root-cause}.

%% file: tex/tables/root_cause_category_stats.tex
\begin{table}[hbpt]
\centering
\caption{Categorization of Numerical Bugs }
\label{table:root_Cause}
\resizebox{0.8\columnwidth}{!}{
 \begin{tabular}{@{}llrrr@{}}
   \toprule
    {\bf Root Cause} & {\bf Sub categories}& \multicolumn{2}{c}{\textbf{Libraries}} & \textbf{Total} \\
\cmidrule(lr){3-4}
 & & \textbf{Closed-Source} & \textbf{Open-Source} &  \\ \midrule

       \multirow{3}{*}{Precision Bugs} & Arithmetic Operation Bugs  & 20   & 11 &  31\\ 
       & Rounding/Approximation Bugs & 25 & 9 &34 \\

       & Casting Bugs & 23& 9& 32\\\hline
       
\multirow{2}{*}{Data Type and Operator Bugs} & Data Type Bugs & 84   & 59 &  143\\ 

 & Operator Bugs & 37 & 57 & 94 \\\hline
 
Special Value Bugs & Special Value Bugs  & 33   & 15 &  48\\ \hline
Device-specific Bugs & Device-specific Bugs& 7   & 8 &  15\\ \bottomrule

    \end{tabular}
}
\end{table}

%% file: tex/tables/symptoms.tex
\begin{table}[hbpt]
\centering
\caption{Categorization of Symptoms }
\label{table:symptoms}

\resizebox{0.8\columnwidth}{!}{
    \begin{tabular}{@{}llrrr@{}}
   \toprule
   {\bf Root Cause} & {\bf Sub categories}& \multicolumn{2}{c}{\textbf{Libraries}} & \textbf{Total} \\\cmidrule(lr){3-4} & & \textbf{Closed-source} & \textbf{Open-source} &  \\ \midrule
            
       \multirow{5}{*}{Error Messages} & Overflow/Underflow    & 21   & 15 & 36\\ 
       
       & Casting & 3 & 8 & 11\\
       
    & Data Type & 57 & 33 & 90 \\
    
    & Operator & 19 & 18 & 37\\
       
       & Undefined Computation & 4 & 3 & 7 \\\hline

\multirow{5}{*}{Wrong Results} & Before and After Casting & 17 & 3    & 20  \\  

& Different Data Types &  9 & 17 & 26 \\

& Different Precision Range & 16 & 34 & 50  \\

& NaN Outputs & 30 & 13 & 43 \\

& Different Amount of Precision & 27 & 13 & 40 \\\hline

\multirow{3}{*}{Performance Bottlenecks}& Cast Operations & 2   & 0  & 2  \\

& Data Types & 17 & 4 & 21 \\

& CPU Programs & 7 & 7 & 14 \\\bottomrule


    \end{tabular}
}
\end{table} %

%% file: tex/bugDetection.tex
\section{Bug Detection Strategies}
\label{sec: manifestation}


Our study highlights numerical bug detection strategies, as summarized in Fig.~\ref{fig:Overview}. Based on findings from Table~\ref{tab:Findings-Implications}, we identify two primary components in detecting GPU-NBs: \textbf{test input generation} and \textbf{test oracles}. The goal of test input generation is to create inputs that can trigger numerical bugs. We explored two methods: \textit{Predefined General Inputs} and \textit{Feedback-Driven Customized Inputs}. Once inputs are generated, the second stage involves applying test oracles to identify GPU-NBs. Test oracles use output comparisons to detect bugs, with some strategies requiring single test runs (e.g., Strategies 1-5 and 9 in Table~\ref{table:ManifestationStrategies}), while others rely on multiple runs. 

\subsection{Test Input Generation} 
\label{subsec: input}

Our findings indicate that many developers rely on random input modifications to trigger numerical bugs in GPU programs. Developers start with predefined inputs and apply random perturbations, such as altering precision, to generate new inputs. This ad hoc approach can be systematically automated using fuzzing techniques, which offer a structured means of input variation to improve bug detection (F-7 and I-7 in Table~\ref{tab:Findings-Implications}).

\subsubsection{Predefined General Inputs}

These inputs serve as the initial test values used to run the program and are categorized into four types: \textit{out-of-range numbers, integers, floating-point numbers}, and \textit{NaN}. Out-of-range numbers fall outside the representable range of a given data type, such as the range for 32-bit floating-point numbers, which spans from 1.17549435E-38 to 3.40282347E+38~\cite{numbers}. NaN values represent undefined or unrepresentable results, often arising from invalid computations or missing inputs ~\cite{NaN}. Floating-point numbers of various precision levels, such as half, single, double, and decimal, are used to test for precision-related issues. Additionally, integers serve as test inputs due to CUDA's implicit conversion of integers to floats, which may introduce precision errors. This categorization aids in understanding how distinct input types impact numerical accuracy in GPU computations.
\begin{figure}[!htbp]
\centering
\includegraphics[width=0.9\columnwidth]{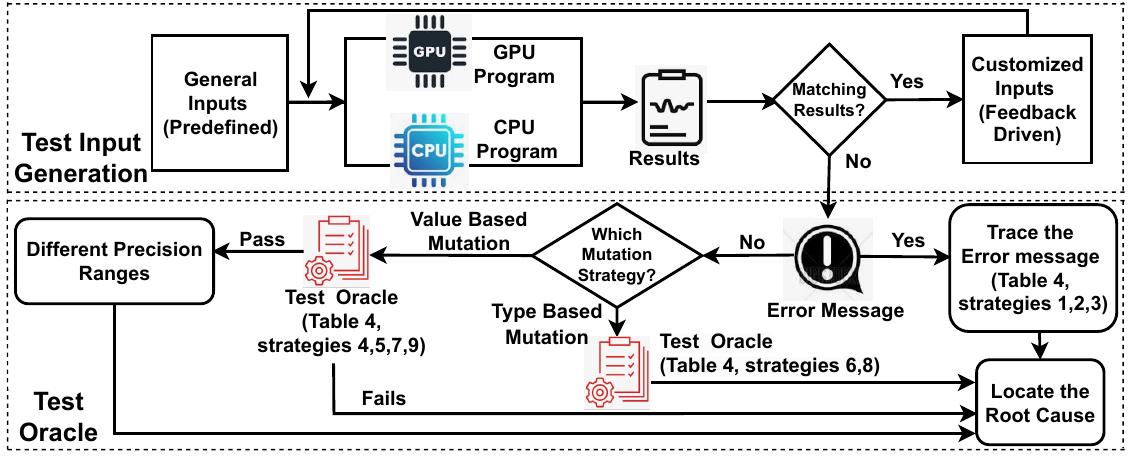}
\caption{Overview on Detecting Numerical Bugs in GPU programs}
\label{fig:Overview}
\end{figure}

\subsubsection{Feedback Driven Customized Inputs}
These inputs leverage mutation techniques to systematically generate inputs that trigger numerical bugs. These inputs are categorized into \textit{value-based} and \textit{type-based} mutations. Value-based mutation employs fuzzing techniques~\cite{greybox} to systematically vary numerical values, increasing the likelihood of identifying GPU-NBs. Type-based mutation \cite{greybox} introduces diverse data types to verify the program's handling of type-specific issues. Advanced methods like Greybox fuzzing \cite{greybox} can be employed to apply these mutations, automating the process of discovering new inputs that might trigger hidden bugs.

\subsection{Test Oracles }
\label{subsec: Oracles}



Test oracles are essential for verifying the correctness of results in GPU programs. As shown in Fig.~\ref{fig:Overview}, our study finds that correctness is initially confirmed by comparing GPU results with outputs from corresponding CPU programs using the same predefined inputs. When discrepancies arise, customized inputs or manifestation strategies can be applied to further investigate numerical bugs (F-8 and I-8 in Table~\ref{tab:Findings-Implications}). These strategies rely on assertion thresholds to differentiate expected numerical variations from valid bugs, and thus require carefully calibrated error bounds to ensure reliable detection.

To define these thresholds, we conducted an empirical evaluation across a diverse set of GPU programs and input types. For each manifestation strategy (Table~\ref{table:ManifestationStrategies}), we measured the distribution of deviations observed between CPU and GPU outputs, or across data type and performance profiles. Thresholds such as 1E-7 for rounding bugs and 0.9 for casting bugs were selected to capture meaningful numerical divergence while minimizing false positives. Similarly, thresholds like 0.01 for data type bugs and 2.5x–3x slowdowns for performance profiling were informed by consistent performance gaps observed across multiple devices and libraries. These values were validated through repeated testing and manual review, ensuring that the selected bounds were robust across architectures and sensitive enough to detect GPU-NBs.

When result discrepancies are observed, error messages offer crucial insight into the root cause, as each numerical bug typically generates a distinctive message. Strategies 1-3 in Table~\ref{table:ManifestationStrategies} provide guidance on implementing exceptions for detecting overflow, underflow, and special value bugs. If no error message is present, alternative manifestation strategies based on the mutation technique used to generate test inputs can be employed.

For value-based mutation, logging and profiling mechanisms can locate numerical bugs. Strategies 4 and 5 in Table~\ref{table:ManifestationStrategies} outline approaches for detecting incorrect results. Rounding bugs, for example, can be identified by comparing logged precision levels between CPU and GPU outputs (e.g., Fig.~\ref{fig: Accuracy-eg}), while casting bugs are detected when pre- and post-cast values differ (e.g., Fig.~\ref{fig: casting-eg}). Profiling can reveal performance bottlenecks, as casting to lower data types may unexpectedly perform slower than higher types (e.g., Fig.~\ref{fig:Ng1} and ~\ref{fig:Ng2}). Device-specific bugs become evident when GPU execution time is slower than CPU execution, as seen in CuPy~\cite{environment1}. If all other strategies yield consistent results, users can test with varying precision ranges to detect potential operator issues.

For type-based mutation, logging and profiling are also effective tools (strategies 6 and 8 in Table~\ref{table:ManifestationStrategies}). Programs are executed with multiple data types on the same input values, requiring at least two runs to enable comparative analysis. Data type bugs are identified when program results diverge significantly beyond minor tolerances (e.g., exceeding a difference of 0.01). Profiling execution times across data types can further reveal data type bugs, especially when operations on a lower precision type take longer than those on a higher precision type.

\input{tex/tables/bugdetection.tex}

\subsection{Preliminary Tool}
\label{subsec: tool}

Based on the findings from Sections~\ref{subsec: input} and~\ref{subsec: Oracles}, we developed GPU-Numerical Bug Detector (GPU-NBDetect) to detect GPU-NBs. GPU-NBDetect integrates automated input generation with differential testing against CPU programs to validate numerical correctness in GPU computations. Its detection logic is guided by the manifestation strategies outlined in Table~\ref{table:ManifestationStrategies} and is informed by key insights from Table~\ref{tab:Findings-Implications}, particularly the importance of input generation (F-7, I-7) and cross-platform output comparison (F-8, I-8) for uncovering GPU-NBs.




GPU-NBDetect uses a template-based approach to automatically generate CUDA and C code for testing. The tool dynamically creates the necessary files in Python, requiring users to simply specify the target mathematical APIs. For CPU programs, the templates include predefined operators, input argument handling, and essential libraries such as \verb|math.h| for numerical operations and exception handling. The GPU templates cover both host and device code, managing memory allocation, data transfer, and kernel execution. Differential testing is performed by executing each compiled program through Python's subprocess interface.




GPU-NBDetect begins testing with user-defined inputs, initialized by selecting random values within the supported range of each mathematical operator, based on documentation~\cite{Math}. To efficiently explore the vast input space and uncover bugs, the tool employs seven mutation operators inspired by fuzzing techniques~\cite{ma2022numfuzz}. These mutators generate diverse numerical inputs, particularly near boundary values (e.g., 0.5, 0.499999, 0.50000001), reflecting the common developer practice of manually probing edge cases to reveal numerical issues. The tool employs the following mutators:


\begin{itemize}[left=0pt]
    \item \textit{bitflip\_s}: Flips the sign component of a floating-point number with a 50\% probability, switching between 0 and 1.
    \item \textit{bitflip\_e}: Flips a single bit at a random position within the exponent.
    \item \textit{bitflip\_m}: Flips a single bit at a random position within the mantissa.
    \item \textit{add\_e}: Adds a randomly generated integer (between 1 and 32) to the exponent.
    \item \textit{add\_m}: Adds an integer \(2^p - 1\) to the mantissa, where p is randomly generated between 1 and 33.
    \item \textit{sub\_e}: Subtracts a randomly generated integer (between 1 and 32) from the exponent. 
    \item \textit{sub\_m}: Subtracts an integer \(2^p - 1\) from the mantissa, where p is randomly generated between 1 and 33.
    \item \textit{random\_bit}: Sets a randomly chosen bit in the exponent or mantissa to a random value. 
\end{itemize}

To detect discrepancies, GPU-NBDetect first applies differential testing by comparing GPU outputs with CPU results. Detected bugs are then classified using assertion-based manifestation strategies (Table~\ref{table:ManifestationStrategies}), which target overflow, underflow, rounding, special values, casting, and data type bugs. To ensure robust and accurate classification, multiple error bounds were empirically tested to establish effective thresholds for each strategy.



\textbf{Minimizing false positives.} To reduce false positives, GPU-NBDetect applies input constraints by (1) restricting values to supported floating-point ranges and (2) excluding known bug-triggering inputs identified in official documentation (e.g., CUDA Math Library~\cite{Math}). Additionally, multiple error bounds were tested to calibrate assertion thresholds for each manifestation strategy (Table~\ref{table:ManifestationStrategies}). Bug samples that failed to consistently meet these thresholds were excluded, improving precision. While this approach ensures accurate bug identification through meaningful result comparisons, profiling-based strategies (strategies 7–9) may still produce occasional false positives due to tool sensitivity and broader input variability.

\input{tex/tables/bugevaluation}

\input{tex/tables/bugevaluation_cupy}

\subsubsection{Evaluation Setup} 

\textcolor{black}{We conducted experiments on two types of GPU programming environments: C-based and Python-based GPU libraries.}
\textcolor{black}{For C-based libraries, we utilized CUDA Math~\cite{Math} and CUDA Thrust~\cite{Thrust}. To benchmark the performance of GPU-NBDetect, we compared it against four established C-based GPU tools: XScope~\cite{laguna2022finding}, which employs Bayesian optimization for input generation; XScope-RS (Random Sampling)~\cite{laguna2022finding}, which utilizes random sampling techniques; BinFPE~\cite{binfpe}, which employs the NVBit dynamic binary instrumentation framework to inspect machine registers for exceptions following each computation; and GPU-FPX~\cite{gpu-fpx}, another NVBit-based tool that detects and analyzes floating-point exceptions on NVIDIA GPUs via binary instrumentation. Since BinFPE~\cite{binfpe} and GPU-FPX~\cite{gpu-fpx} rely on predefined inputs and do not incorporate advanced input generation techniques, we employed a random input generation strategy akin to random sampling to ensure a fair comparative analysis.}

\textcolor{black}{For Python-based GPU libraries, we evaluated CuPy~\cite{cupy} and Numba~\cite{numba}. Notably, GPU-NBDetect is the first approach to systematically test Python-based CUDA math APIs for numerical bugs. Due to the lack of direct baselines for comparison in Python-based GPU environments, we adopted different input generation strategies. Specifically, we leveraged Bayesian optimization (BO) and random sampling (RS)~\cite{laguna2022finding} techniques to evaluate the effectiveness of GPU-NBDetect. } 

\textcolor{black}{To maintain a focus on mathematical computations, we excluded APIs that do not perform numerical operations (e.g., Min, Max) across all evaluated libraries. We tested 53 functions in both CUDA Math and Thrust, and 40 functions in both CuPy and Numba. }
\textcolor{black}{Furthermore, we did not include DL-based numerical bug detection approaches~\cite{DLGPU1, DLGPU4}, as these methods primarily focus on model mutation or hidden-state mutation rather than input mutation. However, XScope-RS employs a strategy similar to the tensor partitioning technique~\cite{DLGPU3}, where input components—mantissa, exponent, and sign—are randomly sampled. The key advantage of GPU-NBDetect lies in its robust input generation, mutation, and manifestation strategies, which enable more comprehensive numerical bug detection across a wide range of GPU libraries. }

Experiments were conducted on a server with an Intel Xeon E5-26 CPU and eight NVIDIA 1080 Ti GPUs, using CUDA version \textcolor{black}{11.1}. For each mathematical function, GPU-NBDetect generates 100 input samples, with three independent test cycles ensuring result consistency and minimizing the randomness from mutation operators.

\subsubsection{Results and Analysis}
\textcolor{black}{Tables~\ref{table:bugdetectionstats} and ~\ref{table:bugdetectionstats_cupy} summarize GPU-NBs identified by GPU-NBDetect, which detected a total of 226 bugs across 186 mathematical functions in four libraries, with 60 confirmed by developers. We measure the average time overhead for bug detection in seconds.}

\textbf{Bug confirmation process.} We reported the identified bugs on the NVIDIA developer forum, where NVIDIA developers reviewed and verified the findings. Although CUDA documentation (e.g., Math Library~\cite{Math}) acknowledges that certain inputs may cause bugs, it does not clarify whether these occur within the supported input range. GPU-NBDetect identified these ambiguous cases, prompting developers to create unit tests for validation. Developers confirmed that GPUs lack built-in exception handling and employ different rounding modes compared to C programs. Notably, GPU-NBDetect pinpointed bug-triggering inputs within the valid input range.

GPU-NBDetect reveals six types of numerical bugs through its analysis. GPU-NBDetect identified rounding errors in 52 functions (27.9\%), highlighting significant precision issues. Developers remarked that \textit{``although the results are within the supported ULP range, these deviations represent errors from the true values.''} Notably, rounding bugs were more stable in CuPy and Numba than in C-based GPU libraries, suggesting that Python-based libraries have more robust rounding modes. Overflow, underflow, and special value bugs were also prevalent but challenging to detect due to the absence of floating-point exception handling in GPUs. As developers clarified, \textit{``CUDA device code does not handle exceptions, so it's essential to verify the behavior against C code.''} Casting bugs were detected in 14 operations, indicating that GPU casting mechanisms are particularly error-prone across both C-based and Python-based libraries. Furthermore, data type bugs emerged when lower precision values led to less efficient calculations compared to higher precision numbers.


In comparison with C-based CUDA libraries and other tools, GPU-NBDetect demonstrated several advantages. GPU-NBDetect detected the highest number of GPU-NBs among all evaluated tools, and all bugs identified by other tools were also successfully detected by GPU-NBDetect, demonstrating its comprehensive coverage. While existing tools, such as XScope and GPU-FPX, produced a significant number of false positives—17 and 33 respectively—GPU-NBDetect minimized false positives by verifying whether the generated inputs fell within the supported numerical range. Furthermore, whereas other tools often generate extreme values, such as 1.3400e+221 and 9.2579e+282, which fall outside the valid range, GPU-NBDetect systematically avoids such unsupported numbers, ensuring more reliable detection. Additionally, unlike existing approaches, GPU-NBDetect successfully detects rounding, casting, and data type bugs, which are not addressed by other tools. Verification is also more challenging with XScope, as it lacks information about the specific inputs that trigger numerical bugs.


\textcolor{black}{ In comparison of input generation techniques in CuPy and Numba}, GPU-NBDetect outperformed RS and BO in detecting GPU-NBs and generating more effective bug-triggering inputs. Unlike RS and BO, which rely on heuristic-based sampling, GPU-NBDetect strategically mutates inputs, leading to a higher bug detection rate across Python-based GPU libraries. These findings highlight the importance of targeted input mutation techniques in uncovering hard-to-detect numerical errors. Futhermore, GPU-NBDetect achieves better overhead efficiency than existing approaches, making it more practical for large-scale GPU applications.

Some of the interesting examples observed by GPU-NBDetect are highlighed: (1) An overflow bug in \textit{cosh} for input 3.6864e+4, exceeding the range specified in the CUDA Math Library, was confirmed against CPU results; (2) An underflow bug in \textit{erfc} with input 6.7076092e+7 produced an output too small for floating-point representation, also verified on the CPU; and (3) A casting bug in \textit{\_\_float2int\_rd} for input 1.101256458240e+11 produced the maximum 32-bit integer, indicating insufficient precision for large conversions.

\textbf{Limitations of GPU-NBDetect.}
GPU-NBDetect has several limitations: (1) GPU-NBDetect only supports float and double data types, limiting its effectiveness for detecting errors in other numerical types. Expanding support for a broader range of data types would improve its effectiveness; (2) The tool has been evaluated on only \textcolor{black}{four libraries}. Future work will expand to include more GPU library APIs, but users can use GPU-NBDetect as a starting point for designing more comprehensive numerical bug detectors; (3) The mutation operators in GPU-NBDetect introduce randomness, which can lead to inconsistent detection results. Employing advanced fuzzing techniques or symbolic execution could improve the robustness of test inputs. Despite these limitations, GPU-NBDetect provides a foundational framework for test input generation and test oracle selection, paving the way for future improvements in GPU-NB detection.

\textbf{Challenges in building tools to detect GPU-NBs.} Developing tools to detect GPU-NBs presents unique challenges: (1) Variability in compute capabilities across GPU architectures complicates tool compatibility, limiting portability; (2) Supporting a broad range of numerical operations and data types is resource-intensive, increasing development and testing complexity; and (3) Effective GPU-NB detection requires specialized expertise in GPU computing, including compiler flags, API management, and memory allocation, which can present a steep learning curve. GPU-NBDetect provides a flexible foundation that addresses these challenges, serving as a basis for future, advanced detection tools. 


\textbf{Other numerical bug detection techniques.} In the future, developers could improve GPU-NBDetect by adapting certain bug detection strategies for GPU-NB detection. For instance, \textit{symbolic execution}~\cite{baldoni2018survey} could be customized for GPU programs to explore diverse execution paths, helping to identify edge cases where numerical bugs may emerge. Similarly, adapting tools like \textit{Clang Static Analyzer}~\cite{lattner2008llvm} to GPU code could enable static analysis for detecting overflow, underflow, and precision errors, though this would require engineering modifications to focus on GPU-specific numerical bugs.
Finally, ~\textit{machine learning models} trained on GPU numerical bug patterns could add an advanced layer of automated detection to GPU-NBDetect, aiding in the early identification of subtle indicators of numerical instability. While each approach would need careful adaptation to suit the GPU domain, integrating these methods could significantly expand the future capabilities of GPU-NBDetect in addressing GPU-specific numerical bugs.

%% file: tex/tables/bugdetection.tex
\begin{table}[hbpt]
\caption{Manifestation Strategies and Assertions to Detect the Root Cause} 
\label{table:ManifestationStrategies}
\centering
\def\arraystretch{}
\resizebox{\columnwidth}{!}{
    \begin{tabular}{|l|p{3cm}|p{5cm}|p{3.8cm}|p{3cm}|}
    \hline
    {\bf }&{\bf Program Iterations (input generations, programs)  }  & {\bf Manifestation Strategy}     & {\bf Assertion} &{\bf Root Cause}\\ \hline
    1 & Single input, single GPU iteration & \textbf{Logging} result A of GPU version & if A $>$ MAX(FP) or A$<$ -MAX(FP) & Overflow bug \\\hline
    2 & Single input, single GPU iteration & \textbf{Logging} result A of GPU version & if A $<$ MIN(FP) or A$>$ -MIN(FP) & Underflow bug \\\hline
    3 & Single input, single GPU iteration & \textbf{Logging} result A of GPU version & if A == NaN  & Special value bug \\\hline
    4& Single input, single CPU and GPU iteration &  \textbf{Logging} precision in result A of CPU version and result B of GPU version & abs(A-B) $>$ 1E-7 & Rounding bug \\\hline

    5 & Single input, single GPU iteration  & \textbf{Logging} value A before casting and value B after casting & abs(B-A) $>$ 0.9  & Casting bug \\\hline 

    6& Multiple inputs, multiple GPU iterations  & \textbf{Logging} result A of first run and result B of another run with different data types for the same value & abs(A- B) $>$ 0.01 & Data type bug \\\hline

    7 & Multiple inputs, multiple GPU iterations &  \textbf{Profiling} execution time A of cast operation and time B of cast operation with higher data type & lower data type A slower than B (avg 2.5X slower) & Casting error bug \\\hline 

    8 & Multiple inputs, multiple GPU iterations &  \textbf{Profiling} overall time A of one data type and time B of higher data type & A is slower than B (nearly 3x slower) & Data type bug \\\hline

    9 & Single input, single CPU and GPU iterations &  \textbf{Profiling} time A of CPU version and time B of GPU version & A is faster than B & Device-specific bug \\\hline

    \end{tabular}
}

\end{table}

%% file: tex/tables/bugevaluation.tex
\begin{table}[hbpt]
\caption{Summary of Detected Numerical Bugs by GPU-NBDetect on C-based GPU Libraries} 
\label{table:bugdetectionstats}

\resizebox{\linewidth}{!}{
\begin{tabular}{|l|l|c|c|c|c|c|c|c|c|}

\hline
  Library & Tool & Overflow Bug &  Underflow Bug & Rounding Bug & Casting Bug & Data Type Bug & Special Value Bug & Total Bugs & Overhead\\
\hline
 CUDA Math & XScope & 8 & 8 & - & - &-  & \textbf{12} & 28 & 8.7\\
                  & XScope-RS & 4 & 5 & - & - & - & 4  & 13 & 2.7\\
                           & BinFPE & 5 & 2 & - & - & - & 5 &  12 & 3.2\\
                            & GPU-FPX & 6 & 4 & - & - & - & 5 & 15 & 4.2\\
                            & GPU-NBDetect & \textbf{10} & \textbf{14} & \textbf{24}  & \textbf{10}  & \textbf{14}  & \textbf{12} & \textbf{84} & \textbf{1.1}\\ \hline
                            
                  CUDA Thrust & XScope & 4  &  7& - & - & - &6  & 17 & 8.4\\
                              & XScope-RS & 4 & 4 & - & - & - & 5  &  13 & 3.4\\
                              & BinFPE & 5 & 6 & - & - & - & 8  &19 & 3.9\\
                              & GPU-FPX & 5 & 8 & - & -  & - & 9  & 22 & 5.0\\
                              & GPU-NBDetect & \textbf{8}  & \textbf{16} & \textbf{20} & \textbf{4} & \textbf{8} &  \textbf{14} & \textbf{70} & \textbf{1.3}\\ \hline

\end{tabular} }

\end{table}

%% file: tex/tables/bugevaluation_cupy.tex
\begin{table}[hbpt]
\caption{Summary of Detected Numerical Bugs by GPU-NBDetect on Python-based GPU Libraries} 
\label{table:bugdetectionstats_cupy}
\resizebox{\linewidth}{!}{
\begin{tabular}{|l|l|c|c|c|c|c|c|c|c|}

\hline
  Library & Tool & Overflow bug &  Underflow bug & Rounding bug & Casting bug & Data Type bug & Special Value bug & Total bugs & Overhead \\
\hline
 
 CuPy & RS & 6 & 4 & 2 & 0 & 2 & 4 & 16& 2.1\\
                            & BO & 7 &4 &2 &0  &2  &4  &  19 & 2.2\\
                            & GPU-NBDetect & \textbf{9} & \textbf{7} & \textbf{4} & \textbf{3} & \textbf{4} & \textbf{8} & \textbf{35} & \textbf{1.7}\\ \hline
                           
                       Numba & RS & 6 & 3 & 2 &1  &3  &4  & 19& 2.3\\
                             & BO & 7 & 5 & 3 &2  &3  & 6 &26 & 2.3\\
                             
                             & GPU-NBDetect &\textbf{9}
                             & \textbf{8} &\textbf{4} & \textbf{3} & \textbf{4} & \textbf{9} & \textbf{37}& \textbf{1.6}\\ \hline

\end{tabular} }

\end{table}

%% file: tex/bugfixing.tex
\section{Fixing Numerical Bugs}
\label{sec: bugfixing}
Our analysis identifies common strategies developers use to manually address GPU-NBs. For each root cause, we examined common fixes based on commit histories, comments, and bug report descriptions. Table~\ref{table:fixing strategies} summarizes these fixing strategies.

\subsection{Fixing Precision Bugs} 


Our study outlines several strategies for addressing precision bugs in GPU programs, as shown in Fig.~\ref{fig:Overviewfinal}. A common approach is adjusting the precision level: increasing precision can improve accuracy for out-of-range values, while decreasing precision may save memory at the cost of precision. Therefore, precision adjustments should be made with careful consideration of application requirements.


Another approach is to implement checkers that identify variables requiring precision adjustments, allowing developers to round values to appropriate levels. For example, issues like those in Fig.~\ref{fig: AccuracyOverflow-eg} can be resolved by casting to higher precision types. Similarly, rounding bugs—such as those caused by computations with 0.5 (Fig.~\ref{fig: Accuracy-eg})—can be mitigated by slight input adjustments (e.g., 0.49999 or 0.50001), as demonstrated in SciPy.


\input{tex/tables/fixingTable} 

Casting bugs can be addressed through upcasting or downcasting. Upcasting converts values to higher precision types, while downcasting reduces precision with rounding to preserve accuracy. For example, when a Decimal64 value cannot be represented in int32 (Fig.~\ref{fig: casting-eg}), casting to float64 provides a correct and effective resolution (see F-15 and I-15 in Table~\ref{tab:Findings-Implications}).

\subsection{Fixing Data Type or Operator Bugs }
Resolving data type and operator bugs in GPU programs often requires a deep understanding of CUDA numerical libraries (F-17 and I-17 in Table~\ref{tab:Findings-Implications}). Data type bugs are typically addressed by adding support for the required data types or substituting with compatible alternatives. Operator bugs may require refactoring numerical operations to ensure accuracy and functional correctness.

The CuPy library uses custom kernel strategies~\cite{user-defined} such as \verb|RawKernel|, \verb|ReductionKernel|, and \verb|ElementwiseKernel| to define operators specifically for GPU programs. These custom kernels, absent in libraries like NumPy and SciPy, are specifically designed to improve numerical stability in CuPy (F-16 and I-16 in Table~\ref{tab:Findings-Implications}). An example of this approach is used to fix the data type bug in Fig.~\ref{fig: Correctness-eg} by modifying a ReductionKernel to support operations like \verb|cupy.mean|, as demonstrated in the CuPy project~\cite{correctness1}. By explicitly defining the mean operation within ReductionKernel, users ensure more accurate computations, illustrating how specialized kernel strategies can address limitations in standard libraries.

\subsection{Fixing Special Value Bugs }
Fixing special value bugs involves adding checks for NaN and INF values. NaN checks identify undefined numbers, while INF checks prevent operations with values smaller than the data type's minimum. For instance, the issue in Fig.~\ref{fig: SpecialValue2-eg} can be fixed by clamping small values that are close to the numerical precision limits of float16, such as rounding values close to zero to exactly zero.

\subsection{Fixing Device-specific Bugs }
Fixing device-specific issues involves two main strategies: (1) switching to a GPU with better numerical support or (2) adjusting data types for mixed precision computations. For the issue in CuPy project~\cite{environment1}, changing the data type to FP32 reduced execution time, with the NumPy version taking 0.255 seconds and the CuPy version taking 0.156 seconds, demonstrating improved performance on the GPU.

%% file: tex/tables/fixingTable.tex
\begin{table}[t]
\centering
\caption{Summary of Fixing Strategies} 
\label{table:fixing strategies}
\resizebox{0.65\linewidth}{!}{
    \begin{tabular}{@{}lrrrr@{}}
   \toprule
        {\bf Fix} & \multicolumn{2}{c}{\textbf{Libraries}} & \textbf{Total} \\ \cmidrule(lr){2-3} &  \textbf{Closed-source} & \textbf{Open-source} &  \\ \midrule

      Fix through arithmetic checkers &  17    & 10   & 27 \\ \hline
      
      Changing Precision &   28   & 10   & 38 \\ \hline
      
      Upcast/Downcast strategies &    23  & 9   & 32 \\ \hline
      
      Refactor Operator Implementation &   39   & 55   & 94 \\ \hline
      
      Custom Kernel Strategy  & 1  & 4 & 5 \\ \hline
      
      \begin{tabular}[c]{@{}l@{}}Add Data Type Support/ \\ Change Data Type\end{tabular}  &84   &59 & 143 \\ \hline

      Fix through NaN checkers     & 31  & 16 & 47 \\ \hline

Use Mixed Precision or Change Type  &6   &5 & 11 \\ \hline

    \end{tabular}
}
\end{table} %

%% file: tex/threat.tex
\section{Threats to Validity}
Our findings are based on a set of bugs we examined, and we applied several filtering steps to ensure correctness and relevance. First, we searched the GitHub database using strict conditions to ensure issues were related to numerical computations in GPU programming. We only selected issues confirmed as real bugs, specifically those marked as closed and containing a bug fix. We verified bug fixes by reviewing the commits associated with each bug report. Keyword filtering identified relevant numerical issues due to the large volume of issues. Each bug was independently inspected by two authors to ensure accuracy. Our results are based on both built-in and third-party numerical libraries in GPU computing, which require special attention compared to CPU computing and other GPU program bugs, enhancing generalizability. 


Regarding the representativeness of our dataset, we focused on repositories associated with numerical issues in GPU programming and restricted our selection to projects impacting real applications. We employed strict criteria in our GitHub search to ensure that the samples were both targeted and representative. For methodological correctness, we collected all available commits from repositories related to numerical issues in GPU programming and utilized the GitHub Archive to find all pertinent numerical issues. We limit the numerical implications of our study to the collected dataset and focus on qualitative insights into the features of the test samples.

%% file: tex/related.tex
\section{Related Work}

\noindent
\textbf{\textcolor{black}{CPU floating point issues vs. GPU floating point issues.}}
Numerous studies have explored floating-point issues in CPU programming~\cite{zou2019detecting, wang2016detecting, zou2015genetic, tang2010perturbing, barr2013automatic, Franco, fu2019effective, DLbug, yi2019efficient}. For instance, Zou et al.~\cite{zou2019detecting} used dynamic analysis to detect program inputs that cause floating-point errors. Wang et al.~\cite{wang2016detecting} proposed a method for detecting and adjusting precision in the GNU C math library. Zou et al.~\cite{zou2015genetic} utilized a search-based approach to generate test inputs that trigger inaccuracies. Tang et al.~\cite{tang2010perturbing} developed a technique for obtaining high-level information about numerical stability and accuracy. Barr et al.~\cite{barr2013automatic} introduced Ariadne to detect floating-point exceptions in the GNU Scientific Library (GSL). Franco et al.~\cite{Franco} conducted a comprehensive study on numerical bugs in real-world numerical libraries.

Several studies have addressed floating-point issues in GPU programming. Laguna ~\cite{laguna2019fpchecker, binfpe, laguna2022finding} introduced tools like FPCHECKER, BinFPE, and XScope to automatically detect floating-point exceptions for FP32 and FP64 data types in GPU applications. Our study focuses on numerical bugs within the supported range, addressing precision issues for various numerical operators in CUDA libraries and different data types. We incorporate differential testing to verify results against CPU outputs and use seven mutation operators for test input generation. Surineni et al.~\cite{surineni2017understanding} examined performance-accuracy trade-offs in different arithmetic precision. Polok et al.~\cite{polok2016increasing} studied increasing double precision throughput on Maxwell GPUs. Ho et al.~\cite{ho2017exploiting, ho2021gram} developed frameworks for exploiting Pascal's 
half-precision capabilities and dynamically mixing precision in GPU applications. While prior studies have analyzed GPU programming bugs ~\cite{li2019detecting,tiwari2015understanding,yang2012fixing111,  wu2020simulee,sorensen2016exposing, islam2018bugaroo,jiang2020cudasmith, gopalakrishnan2021guarding,DLGPU7}, they often focus on synchronization and memory issues rather than numerical bugs.

\noindent
\textcolor{black}{\textbf{DL-based numerical issues vs. GPU-NBs.}}
\textcolor{black}{Prior research on numerical issues in DL has explored how flaws in algorithm design and logic, including unstable optimization methods, poorly conditioned loss functions, and incorrect gradient computations, give rise to application-level numerical instabilities~\cite{DLGPU1,DLGPU2,DLGPU3,DLGPU4,DLbug,DLGPU6,DLGPU7}. Additionally, these studies examine the role of framework-specific transformations in exacerbating these instabilities. These issues manifest as observable deviations in model outputs, making them detectable through accuracy loss or performance degradation.} 
\textcolor{black}{In contrast, our work focuses on GPU-NBs, which arise from how data types and numerical operations are handled during execution and library implementations. These include floating-point precision limits, rounding/truncation errors, and data type mismatches. Unlike application-level issues, which impact application's behavior, GPU-NBs silently introduce errors without explicit failure signals, affecting computations across scientific computing, high-performance computing (HPC), and DL. These issues cannot be detected through application-level techniques, necessitating low-level numerical bug detection techniques that operate independently of algorithmic design.}

\textcolor{black}{Detection methodologies for numerical bugs differ between DL-based approaches and GPU low-level detection techniques. DL-specific techniques, such as metamorphic and differential testing~\cite{DLGPU1,DLGPU3} focus on perturbing architectures and hyperparameters to identify inconsistencies at the model level. However, these methods rely on high-level behavioral deviations, making them ineffective at exposing low-level numerical errors in GPU computations. GPU-NBs are inherently stealthy, often remaining undetected at the model's output since their effects may not consistently propagate through multiple layers of computation.}

\textcolor{black}{A major limitation of DL-based approaches is the immense search space required to expose numerical bugs through model perturbation~\cite{DLGPU1,DLGPU3}. Identifying arithmetic errors this way would necessitate exhaustive exploration of architectures and parameter configurations to find specific conditions where these errors amplify into observable inconsistencies. Given the combinatorial complexity of DL models, this approach is both inefficient and unreliable for detecting GPU-NBs.}

\textcolor{black}{In contrast, GPU-NBDetect employs a direct, low-level detection mechanism, systematically modifying floating-point precision and rounding behavior to uncover numerical inconsistencies at their source. By leveraging fine-grained bit-level perturbations and cross-device validation, GPU-NBDetect detects numerical flaws independent of model structure, addressing a fundamental gap in DL-based methodologies. The inability of DL-based testing to explicitly target numerical operations underscores the necessity of a dedicated low-level detection framework for GPU-NBs.}

\textcolor{black}{Device-specific numerical bugs are a common type of numerical issue, but their handling differs significantly between DL-focused studies and our approach. DL studies often attribute these bugs to library version mismatches, recommending updates as a solution~\cite{DLGPU2,DLGPU4, DLGPU6}. However, this approach overlooks architectural constraints in certain GPU models, which may lack support for specific numerical operations or precision levels. Our study highlights the need for precision-aware computation strategies and effective data type selection, particularly for GPUs with limited numerical capabilities. By systematically analyzing these issues, we provide practical guidance on mitigating hardware-specific numerical errors, addressing a gap in prior research.}

\textcolor{black}{Beyond DL, GPU-NBs impact a wide range of GPU-based applications, including scientific computing and high-performance computing (HPC). Unlike DL numerical bugs, which primarily affect model accuracy and stability, GPU-NBs introduce fundamental computational errors that can compromise scientific results and numerical reliability in mission-critical applications. This study is the first to systematically investigate GPU-NBs at the arithmetic level. By introducing GPU-NBDetect, we provide a generalizable framework for detecting numerical inconsistencies across all GPU applications, rather than focusing solely on DL optimizations. Our findings emphasize the need for numerical bug detection techniques to ensure robust and reliable GPU-based computations.}

%% file: tex/conclusion.tex
\section{Conclusion}

This work presents a systematic investigation into GPU-NBs, offering a fine-grained characterization of their root causes, symptoms, and repair strategies. Through an empirical analysis of 397 validated bug reports from both open-source and closed-source libraries, we classify these issues into distinct categories and uncover their underlying bug patterns. Building on these insights, we propose GPU-NBDetect, a preliminary detection tool that integrates mutation-based input generation and oracle-driven validation to identify a wide range of GPU-NBs. GPU-NBDetect demonstrates strong detection capability across diverse libraries, providing a framework for automated analysis. The findings and tools introduced in this study lay the groundwork for future research on testing and repairing numerical software in GPU-based systems.

\section*{Data Availability}
Our dataset and project website are available at ~\cite{GPUstudy}.

\section*{Acknowledgments}
This work was partially supported by NSF grants NSF CCF-2146443 and Amazon Research Award, Fall 2024.




%% file: main.bbl

\begin{thebibliography}{84}


\ifx \showCODEN    \undefined \def \showCODEN     #1{\unskip}     \fi
\ifx \showISBNx    \undefined \def \showISBNx     #1{\unskip}     \fi
\ifx \showISBNxiii \undefined \def \showISBNxiii  #1{\unskip}     \fi
\ifx \showISSN     \undefined \def \showISSN      #1{\unskip}     \fi
\ifx \showLCCN     \undefined \def \showLCCN      #1{\unskip}     \fi
\ifx \shownote     \undefined \def \shownote      #1{#1}          \fi
\ifx \showarticletitle \undefined \def \showarticletitle #1{#1}   \fi
\ifx \showURL      \undefined \def \showURL       {\relax}        \fi
\providecommand\bibfield[2]{#2}
\providecommand\bibinfo[2]{#2}
\providecommand\natexlab[1]{#1}
\providecommand\showeprint[2][]{arXiv:#2}

\bibitem[ARM(2024)]%
        {NaN}
\bibfield{author}{\bibinfo{person}{ARM}.} \bibinfo{year}{2024}\natexlab{}.
\newblock \bibinfo{title}{ARM architecture reference manual ARMv7-A and ARMv7-R
  edition}.
\newblock
\urldef\tempurl%
\url{https://developer.arm.com/documentation/ddi0406/c/Application-Level-Architecture/Application-Level-Programmers--Model/Floating-point-data-types-and-arithmetic/NaN-handling-and-the-Default-NaN}
\showURL{%
\tempurl}


\bibitem[Baldoni et~al\mbox{.}(2018)]%
        {baldoni2018survey}
\bibfield{author}{\bibinfo{person}{Roberto Baldoni}, \bibinfo{person}{Emilio
  Coppa}, \bibinfo{person}{Daniele~Cono D’Elia}, \bibinfo{person}{Camil
  Demetrescu}, {and} \bibinfo{person}{Irene Finocchi}.}
  \bibinfo{year}{2018}\natexlab{}.
\newblock \showarticletitle{A survey of symbolic execution techniques}.
\newblock \bibinfo{journal}{\emph{Comput. Surveys}}  \bibinfo{volume}{51}
  (\bibinfo{year}{2018}), \bibinfo{pages}{1--39}.
\newblock
\href{https://doi.org/10.1145/3182657}{doi:\nolinkurl{10.1145/3182657}}


\bibitem[Barr et~al\mbox{.}(2013)]%
        {barr2013automatic}
\bibfield{author}{\bibinfo{person}{Earl~T. Barr}, \bibinfo{person}{Thanh Vo},
  \bibinfo{person}{Vu Le}, {and} \bibinfo{person}{Zhendong Su}.}
  \bibinfo{year}{2013}\natexlab{}.
\newblock \showarticletitle{Automatic detection of floating-point exceptions}.
  In \bibinfo{booktitle}{\emph{Proceedings of the 40th ACM SIGPLAN-SIGACT
  Symposium on Principles of Programming Languages (POPL)}}.
  \bibinfo{pages}{549--560}.
\newblock
\href{https://doi.org/10.1145/2429069.2429133}{doi:\nolinkurl{10.1145/2429069.2429133}}


\bibitem[Betts et~al\mbox{.}(2012)]%
        {gpuverify}
\bibfield{author}{\bibinfo{person}{Adam Betts}, \bibinfo{person}{Nathan Chong},
  \bibinfo{person}{Alastair Donaldson}, \bibinfo{person}{Shaz Qadeer}, {and}
  \bibinfo{person}{Paul Thomson}.} \bibinfo{year}{2012}\natexlab{}.
\newblock \showarticletitle{GPUVerify: A verifier for GPU kernels}. In
  \bibinfo{booktitle}{\emph{Proceedings of the ACM International Conference on
  Object Oriented Programming, Systems, Languages, and Applications (OOPSLA)}}.
  \bibinfo{pages}{113--132}.
\newblock
\href{https://doi.org/10.1145/2384616.2384625}{doi:\nolinkurl{10.1145/2384616.2384625}}


\bibitem[Bhatnagar(2008)]%
        {datatypedefinition}
\bibfield{author}{\bibinfo{person}{Seema Bhatnagar}.}
  \bibinfo{year}{2008}\natexlab{}.
\newblock \bibinfo{booktitle}{\emph{Textbook of computer science for class
  XI}}.
\newblock \bibinfo{publisher}{PHI Learning Pvt. Ltd.} 182 pages.
\newblock
\showISBNx{978-81-203-2993-5}
\urldef\tempurl%
\url{https://books.google.com/books?id=bjE5EHw35DkC}
\showURL{%
\tempurl}


\bibitem[Boyer et~al\mbox{.}(2008)]%
        {boyer2008automated}
\bibfield{author}{\bibinfo{person}{Michael Boyer}, \bibinfo{person}{Kevin
  Skadron}, {and} \bibinfo{person}{Westley Weimer}.}
  \bibinfo{year}{2008}\natexlab{}.
\newblock \showarticletitle{Automated dynamic analysis of CUDA programs}. In
  \bibinfo{booktitle}{\emph{Proceedings of the Third Workshop on Software Tools
  for MultiCore Systems}}. \bibinfo{pages}{33}.
\newblock


\bibitem[Committee(2019)]%
        {wikipedia_2022}
\bibfield{author}{\bibinfo{person}{Microprocessor~Standards Committee}.}
  \bibinfo{year}{2019}\natexlab{}.
\newblock \showarticletitle{IEEE standard for floating-point arithmetic}.
\newblock \bibinfo{journal}{\emph{IEEE Std 754-2019 (Revision of IEEE
  754-2008)}} (\bibinfo{year}{2019}), \bibinfo{pages}{1--84}.
\newblock
\href{https://doi.org/10.1109/IEEESTD.2019.8766229}{doi:\nolinkurl{10.1109/IEEESTD.2019.8766229}}


\bibitem[Di~Franco et~al\mbox{.}(2017)]%
        {Franco}
\bibfield{author}{\bibinfo{person}{Anthony Di~Franco}, \bibinfo{person}{Hui
  Guo}, {and} \bibinfo{person}{Cindy Rubio-Gonz{\'a}lez}.}
  \bibinfo{year}{2017}\natexlab{}.
\newblock \showarticletitle{A comprehensive study of real-world numerical bug
  characteristics}. In \bibinfo{booktitle}{\emph{Proceedings of the 32nd
  IEEE/ACM International Conference on Automated Software Engineering (ASE)}}.
  \bibinfo{pages}{509--519}.
\newblock
\href{https://doi.org/10.1109/ASE.2017.8115662}{doi:\nolinkurl{10.1109/ASE.2017.8115662}}


\bibitem[Eizenberg et~al\mbox{.}(2017)]%
        {barracuda}
\bibfield{author}{\bibinfo{person}{Ariel Eizenberg}, \bibinfo{person}{Yuanfeng
  Peng}, \bibinfo{person}{Toma Pigli}, \bibinfo{person}{William Mansky}, {and}
  \bibinfo{person}{Joseph Devietti}.} \bibinfo{year}{2017}\natexlab{}.
\newblock \showarticletitle{BARRACUDA: Binary-level analysis of runtime races
  in CUDA programs}. In \bibinfo{booktitle}{\emph{Proceedings of the 38th ACM
  SIGPLAN Conference on Programming Language Design and Implementation
  (PLDI)}}. \bibinfo{pages}{126--140}.
\newblock
\href{https://doi.org/10.1145/3062341.3062342}{doi:\nolinkurl{10.1145/3062341.3062342}}


\bibitem[Forums(2012)]%
        {person_2012}
\bibfield{author}{\bibinfo{person}{NVIDIA~Developer Forums}.}
  \bibinfo{year}{2012}\natexlab{}.
\newblock \bibinfo{title}{Floating point operations difference between CPU and
  GPU}.
\newblock
\newblock
\shownote{\url{https://forums.developer.nvidia.com/t/floating-point-operations-difference-between-cpu-and-gpu/27334}}.


\bibitem[Fu and Su(2019)]%
        {fu2019effective}
\bibfield{author}{\bibinfo{person}{Zhoulai Fu} {and} \bibinfo{person}{Zhendong
  Su}.} \bibinfo{year}{2019}\natexlab{}.
\newblock \showarticletitle{Effective floating-point analysis via weak-distance
  minimization}. In \bibinfo{booktitle}{\emph{Proceedings of the 40th ACM
  SIGPLAN Conference on Programming Language Design and Implementation
  (PLDI)}}. \bibinfo{pages}{439--452}.
\newblock
\href{https://doi.org/10.1145/3314221.3314632}{doi:\nolinkurl{10.1145/3314221.3314632}}


\bibitem[Ginkgo-Project(2021)]%
        {ginkgo}
\bibfield{author}{\bibinfo{person}{Ginkgo-Project}.}
  \bibinfo{year}{2021}\natexlab{}.
\newblock \bibinfo{title}{Ginkgo: Numerical linear algebra software package}.
\newblock
\urldef\tempurl%
\url{https://github.com/ginkgo-project/ginkgo}
\showURL{%
\tempurl}


\bibitem[GitHub(2021)]%
        {gharchive}
\bibfield{author}{\bibinfo{person}{GitHub}.} \bibinfo{year}{2021}\natexlab{}.
\newblock \bibinfo{title}{GitHub Archive}.
\newblock
\urldef\tempurl%
\url{https://archiveprogram.github.com/}
\showURL{%
\tempurl}


\bibitem[Google(2021a)]%
        {tensorflowbackground1}
\bibfield{author}{\bibinfo{person}{Google}.} \bibinfo{year}{2021}\natexlab{a}.
\newblock \bibinfo{title}{TensorFlow: TensorFlow 2.2 using \texttt{tf.float16}
  executes only on CPU (issue \#41783)}.
\newblock
\urldef\tempurl%
\url{https://github.com/tensorflow/tensorflow/issues/41783}
\showURL{%
\tempurl}


\bibitem[Google(2021b)]%
        {tensorflowbackground3}
\bibfield{author}{\bibinfo{person}{Google}.} \bibinfo{year}{2021}\natexlab{b}.
\newblock \bibinfo{title}{TensorFlow: TFLite \texttt{conv\_2D} unsupported data
  type for float32 tensor (issue \#40357)}.
\newblock
\urldef\tempurl%
\url{https://github.com/tensorflow/tensorflow/issues/40357}
\showURL{%
\tempurl}


\bibitem[Google(2023)]%
        {tensorflowbackground2}
\bibfield{author}{\bibinfo{person}{Google}.} \bibinfo{year}{2023}\natexlab{}.
\newblock \bibinfo{title}{TensorFlow: Mixed precision}.
\newblock
\urldef\tempurl%
\url{https://www.tensorflow.org/guide/mixed_precision#overview}
\showURL{%
\tempurl}


\bibitem[Gopalakrishnan et~al\mbox{.}(2021)]%
        {gopalakrishnan2021guarding}
\bibfield{author}{\bibinfo{person}{Ganesh Gopalakrishnan},
  \bibinfo{person}{Ignacio Laguna}, \bibinfo{person}{Ang Li},
  \bibinfo{person}{Pavel Panchekha}, \bibinfo{person}{Cindy
  Rubio-Gonz{\'a}lez}, {and} \bibinfo{person}{Zachary Tatlock}.}
  \bibinfo{year}{2021}\natexlab{}.
\newblock \showarticletitle{Guarding numerics amidst rising heterogeneity}. In
  \bibinfo{booktitle}{\emph{Proceedings of the 5th IEEE/ACM International
  Workshop on Software Correctness for HPC Applications (Correctness)}}.
  \bibinfo{pages}{9--15}.
\newblock
\href{https://doi.org/10.1109/Correctness54621.2021.00007}{doi:\nolinkurl{10.1109/Correctness54621.2021.00007}}


\bibitem[GPUNB(2024)]%
        {GPUstudy}
\bibfield{author}{\bibinfo{person}{GPUNB}.} \bibinfo{year}{2024}\natexlab{}.
\newblock \bibinfo{title}{GPU program numerical bug study}.
\newblock
\urldef\tempurl%
\url{https://gpu-program-bug-study.github.io/Comprehensive-Study-on-GPU-Program-Numerical-Issues.github.io/}
\showURL{%
\tempurl}


\bibitem[Hales(2018)]%
        {Hales_2018}
\bibfield{author}{\bibinfo{person}{Jeff Hales}.}
  \bibinfo{year}{2018}\natexlab{}.
\newblock \bibinfo{title}{Deep learning framework power scores 2018}.
\newblock
\urldef\tempurl%
\url{https://www.kaggle.com/code/discdiver/deep-learning-framework-power-scores-2018}
\showURL{%
\tempurl}


\bibitem[Ho et~al\mbox{.}(2021)]%
        {ho2021gram}
\bibfield{author}{\bibinfo{person}{Nhut-Minh Ho}, \bibinfo{person}{Himeshi
  De~Silva}, {and} \bibinfo{person}{Weng-Fai Wong}.}
  \bibinfo{year}{2021}\natexlab{}.
\newblock \showarticletitle{GRAM: A framework for dynamically mixing precisions
  in GPU applications}.
\newblock \bibinfo{journal}{\emph{ACM Transactions on Architecture and Code
  Optimization (TACO)}}  \bibinfo{volume}{18} (\bibinfo{year}{2021}),
  \bibinfo{pages}{1--24}.
\newblock
\href{https://doi.org/10.1145/3441830}{doi:\nolinkurl{10.1145/3441830}}


\bibitem[Ho and Wong(2017)]%
        {ho2017exploiting}
\bibfield{author}{\bibinfo{person}{Nhut-Minh Ho} {and}
  \bibinfo{person}{Weng-Fai Wong}.} \bibinfo{year}{2017}\natexlab{}.
\newblock \showarticletitle{Exploiting half precision arithmetic in NVIDIA
  GPUs}. In \bibinfo{booktitle}{\emph{Proceedings of the IEEE High Performance
  Extreme Computing Conference (HPEC)}}. \bibinfo{pages}{1--7}.
\newblock
\href{https://doi.org/10.1109/HPEC.2017.8091072}{doi:\nolinkurl{10.1109/HPEC.2017.8091072}}


\bibitem[IBM(2021)]%
        {numbers}
\bibfield{author}{\bibinfo{person}{IBM}.} \bibinfo{year}{2021}\natexlab{}.
\newblock \bibinfo{title}{Numbers}.
\newblock
\newblock
\shownote{\url{https://www.ibm.com/docs/en/idr/11.4.0?topic=types-numbers}}.


\bibitem[Islam and Muzahid(2018)]%
        {islam2018bugaroo}
\bibfield{author}{\bibinfo{person}{Mohammad~Majharul Islam} {and}
  \bibinfo{person}{Abdullah Muzahid}.} \bibinfo{year}{2018}\natexlab{}.
\newblock \showarticletitle{Bugaroo: Exposing memory model bugs in many-core
  systems}. In \bibinfo{booktitle}{\emph{Proceedings of the 29th IEEE
  International Symposium on Software Reliability Engineering (ISSRE)}}.
  \bibinfo{pages}{178--188}.
\newblock
\href{https://doi.org/10.1109/ISSRE.2018.00028}{doi:\nolinkurl{10.1109/ISSRE.2018.00028}}


\bibitem[Jiang et~al\mbox{.}(2020)]%
        {jiang2020cudasmith}
\bibfield{author}{\bibinfo{person}{Bo Jiang}, \bibinfo{person}{Xiaoyan Wang},
  \bibinfo{person}{Wing~Kwong Chan}, \bibinfo{person}{T.~H. Tse},
  \bibinfo{person}{Na Li}, \bibinfo{person}{Yongfeng Yin}, {and}
  \bibinfo{person}{Zhenyu Zhang}.} \bibinfo{year}{2020}\natexlab{}.
\newblock \showarticletitle{CUDAsmith: A fuzzer for CUDA compilers}. In
  \bibinfo{booktitle}{\emph{Proceedings of the 44th IEEE Annual Computers,
  Software, and Applications Conference (COMPSAC)}}. \bibinfo{pages}{861--871}.
\newblock
\href{https://doi.org/10.1109/COMPSAC48688.2020.0-156}{doi:\nolinkurl{10.1109/COMPSAC48688.2020.0-156}}


\bibitem[Kloberdanz et~al\mbox{.}(2022)]%
        {DLGPU7}
\bibfield{author}{\bibinfo{person}{Eliska Kloberdanz}, \bibinfo{person}{Kyle~G
  Kloberdanz}, {and} \bibinfo{person}{Wei Le}.}
  \bibinfo{year}{2022}\natexlab{}.
\newblock \showarticletitle{DeepStability: A study of unstable numerical
  methods and their solutions in deep learning}. In
  \bibinfo{booktitle}{\emph{Proceedings of the 44th IEEE/ACM International
  Conference on Software Engineering (ICSE)}}. \bibinfo{pages}{586--597}.
\newblock
\href{https://doi.org/10.1145/3510003.3510095}{doi:\nolinkurl{10.1145/3510003.3510095}}


\bibitem[Laguna(2019)]%
        {laguna2019fpchecker}
\bibfield{author}{\bibinfo{person}{Ignacio Laguna}.}
  \bibinfo{year}{2019}\natexlab{}.
\newblock \showarticletitle{FpChecker: Detecting floating-point exceptions in
  GPU applications}. In \bibinfo{booktitle}{\emph{Proceedings of the 34th
  IEEE/ACM International Conference on Automated Software Engineering (ASE)}}.
  \bibinfo{pages}{1126--1129}.
\newblock
\href{https://doi.org/10.1109/ASE.2019.00118}{doi:\nolinkurl{10.1109/ASE.2019.00118}}


\bibitem[Laguna and Gopalakrishnan(2022)]%
        {laguna2022finding}
\bibfield{author}{\bibinfo{person}{Ignacio Laguna} {and}
  \bibinfo{person}{Ganesh Gopalakrishnan}.} \bibinfo{year}{2022}\natexlab{}.
\newblock \showarticletitle{Finding inputs that trigger floating-point
  exceptions in GPUs via Bayesian optimization}. In
  \bibinfo{booktitle}{\emph{Proceedings of the IEEE/ACM International
  Conference for High Performance Computing, Networking, Storage and Analysis
  (SC)}}. \bibinfo{pages}{1--14}.
\newblock
\href{https://doi.org/10.1109/SC41404.2022.00038}{doi:\nolinkurl{10.1109/SC41404.2022.00038}}


\bibitem[Laguna et~al\mbox{.}(2022)]%
        {binfpe}
\bibfield{author}{\bibinfo{person}{Ignacio Laguna}, \bibinfo{person}{Xinyi Li},
  {and} \bibinfo{person}{Ganesh Gopalakrishnan}.}
  \bibinfo{year}{2022}\natexlab{}.
\newblock \showarticletitle{BinFPE: Accurate floating-point exception detection
  for GPU applications}. In \bibinfo{booktitle}{\emph{Proceedings of the 11th
  ACM SIGPLAN International Workshop on the State Of the Art in Program
  Analysis (SOAP)}}. \bibinfo{pages}{1--8}.
\newblock
\href{https://doi.org/10.1145/3520313.3534655}{doi:\nolinkurl{10.1145/3520313.3534655}}


\bibitem[Lattner(2008)]%
        {lattner2008llvm}
\bibfield{author}{\bibinfo{person}{Chris Lattner}.}
  \bibinfo{year}{2008}\natexlab{}.
\newblock \showarticletitle{LLVM and Clang: Next generation compiler
  technology}. In \bibinfo{booktitle}{\emph{Proceedings of the BSD
  Conference}}, Vol.~\bibinfo{volume}{5}. \bibinfo{pages}{1--20}.
\newblock


\bibitem[Li et~al\mbox{.}(2012)]%
        {gklee}
\bibfield{author}{\bibinfo{person}{Guodong Li}, \bibinfo{person}{Peng Li},
  \bibinfo{person}{Geof Sawaya}, \bibinfo{person}{Ganesh Gopalakrishnan},
  \bibinfo{person}{Indradeep Ghosh}, {and} \bibinfo{person}{Sreeranga~P.
  Rajan}.} \bibinfo{year}{2012}\natexlab{}.
\newblock \showarticletitle{GKLEE: Concolic verification and test generation
  for GPUs}.
\newblock \bibinfo{journal}{\emph{ACM SIGPLAN Notices}}  \bibinfo{volume}{47}
  (\bibinfo{year}{2012}), \bibinfo{pages}{215--224}.
\newblock
\href{https://doi.org/10.1145/2370036.2145844}{doi:\nolinkurl{10.1145/2370036.2145844}}


\bibitem[Li et~al\mbox{.}(2019)]%
        {li2019detecting}
\bibfield{author}{\bibinfo{person}{Wentao Li}, \bibinfo{person}{Jianhua Sun},
  {and} \bibinfo{person}{Hao Chen}.} \bibinfo{year}{2019}\natexlab{}.
\newblock \showarticletitle{Detecting undefined behaviors in CUDA C}.
\newblock \bibinfo{journal}{\emph{IEEE Access}}  \bibinfo{volume}{7}
  (\bibinfo{year}{2019}), \bibinfo{pages}{182559--182572}.
\newblock
\href{https://doi.org/10.1109/ACCESS.2019.2954143}{doi:\nolinkurl{10.1109/ACCESS.2019.2954143}}


\bibitem[Li et~al\mbox{.}(2023)]%
        {gpu-fpx}
\bibfield{author}{\bibinfo{person}{Xinyi Li}, \bibinfo{person}{Ignacio Laguna},
  \bibinfo{person}{Bo Fang}, \bibinfo{person}{Katarzyna Swirydowicz},
  \bibinfo{person}{Ang Li}, {and} \bibinfo{person}{Ganesh Gopalakrishnan}.}
  \bibinfo{year}{2023}\natexlab{}.
\newblock \showarticletitle{Design and evaluation of GPU-FPX: A low-overhead
  tool for floating-point exception detection in NVIDIA GPUs}. In
  \bibinfo{booktitle}{\emph{Proceedings of the 32nd ACM International Symposium
  on High-Performance Parallel and Distributed Computing (HPDC)}}.
  \bibinfo{pages}{59--71}.
\newblock
\href{https://doi.org/10.1145/3588195.3592991}{doi:\nolinkurl{10.1145/3588195.3592991}}


\bibitem[Ma et~al\mbox{.}(2022)]%
        {ma2022numfuzz}
\bibfield{author}{\bibinfo{person}{Chenghu Ma}, \bibinfo{person}{Liqian Chen},
  \bibinfo{person}{Xin Yi}, \bibinfo{person}{Guangsheng Fan}, {and}
  \bibinfo{person}{Ji Wang}.} \bibinfo{year}{2022}\natexlab{}.
\newblock \showarticletitle{NuMFUZZ: A floating-point format aware fuzzer for
  numerical programs}. In \bibinfo{booktitle}{\emph{Proceedings of the 29th
  IEEE Asia-Pacific Software Engineering Conference (APSEC)}}.
  \bibinfo{pages}{338--347}.
\newblock
\href{https://doi.org/10.1109/APSEC57359.2022.00046}{doi:\nolinkurl{10.1109/APSEC57359.2022.00046}}


\bibitem[Ma et~al\mbox{.}(2025)]%
        {ma2024symbolic}
\bibfield{author}{\bibinfo{person}{Dongyu Ma}, \bibinfo{person}{Zeyu Liang},
  \bibinfo{person}{Luming Yin}, {and} \bibinfo{person}{Hongliang Liang}.}
  \bibinfo{year}{2025}\natexlab{}.
\newblock \showarticletitle{Symbolic testing of floating-point bugs and
  exceptions}.
\newblock \bibinfo{journal}{\emph{Journal of Systems and Software}}
  \bibinfo{volume}{219} (\bibinfo{year}{2025}).
\newblock
\href{https://doi.org/10.1016/j.jss.2024.112226}{doi:\nolinkurl{10.1016/j.jss.2024.112226}}


\bibitem[Muller et~al\mbox{.}(2018)]%
        {substractive}
\bibfield{author}{\bibinfo{person}{Jean-Michel Muller},
  \bibinfo{person}{Nicolas Brunie}, \bibinfo{person}{Florent de Dinechin},
  \bibinfo{person}{Claude-Pierre Jeannerod}, \bibinfo{person}{Mioara Joldes},
  \bibinfo{person}{Vincent Lefèvre}, \bibinfo{person}{Guillaume Melquiond},
  \bibinfo{person}{Nathalie Revol}, {and} \bibinfo{person}{Serge Torres}.}
  \bibinfo{year}{2018}\natexlab{}.
\newblock \bibinfo{booktitle}{\emph{Handbook of Floating-Point Arithmetic}
  (\bibinfo{edition}{2} ed.)}.
\newblock \bibinfo{publisher}{Birkhäuser}, \bibinfo{address}{Gewerbestrasse
  11, 6330 Cham, Switzerland}. 102 pages.
\newblock
\showISBNx{978-3-319-76525-9}
\href{https://doi.org/10.1007/978-3-319-76526-6}{doi:\nolinkurl{10.1007/978-3-319-76526-6}}


\bibitem[NVIDIA(2009)]%
        {fermi}
\bibfield{author}{\bibinfo{person}{NVIDIA}.} \bibinfo{year}{2009}\natexlab{}.
\newblock \bibinfo{title}{NVIDIA's next generation CUDA compute architecture:
  Fermi v1.1}.
\newblock
\urldef\tempurl%
\url{https://www.nvidia.com/content/PDF/fermi_white_papers/NVIDIA_Fermi_Compute_Architecture_Whitepaper.pdf}
\showURL{%
\tempurl}


\bibitem[NVIDIA(2012)]%
        {kepler}
\bibfield{author}{\bibinfo{person}{NVIDIA}.} \bibinfo{year}{2012}\natexlab{}.
\newblock \bibinfo{title}{NVIDIA's next generation CUDA compute architecture:
  Kepler GK110/210 v1.1}.
\newblock
\urldef\tempurl%
\url{https://www.nvidia.com/content/dam/en-zz/Solutions/Data-Center/tesla-product-literature/NVIDIA-Kepler-GK110-GK210-Architecture-Whitepaper.pdf}
\showURL{%
\tempurl}


\bibitem[NVIDIA(2019a)]%
        {amgx}
\bibfield{author}{\bibinfo{person}{NVIDIA}.} \bibinfo{year}{2019}\natexlab{a}.
\newblock \bibinfo{title}{AMGX: Algebraic multigrid solver (AmgX) library}.
\newblock
\urldef\tempurl%
\url{https://developer.nvidia.com/amgx}
\showURL{%
\tempurl}


\bibitem[NVIDIA(2019b)]%
        {Math}
\bibfield{author}{\bibinfo{person}{NVIDIA}.} \bibinfo{year}{2019}\natexlab{b}.
\newblock \bibinfo{title}{CUDA math library}.
\newblock
\urldef\tempurl%
\url{https://developer.nvidia.com/cuda-math-library}
\showURL{%
\tempurl}


\bibitem[NVIDIA(2021a)]%
        {cublas}
\bibfield{author}{\bibinfo{person}{NVIDIA}.} \bibinfo{year}{2021}\natexlab{a}.
\newblock \bibinfo{title}{cuBLAS: Basic linear algebra on NVIDIA GPUs}.
\newblock
\newblock
\shownote{\url{https://developer.nvidia.com/cublas}}.


\bibitem[NVIDIA(2021b)]%
        {cufft}
\bibfield{author}{\bibinfo{person}{NVIDIA}.} \bibinfo{year}{2021}\natexlab{b}.
\newblock \bibinfo{title}{cuFFT: CUDA fast Fourier transform library}.
\newblock
\urldef\tempurl%
\url{https://developer.nvidia.com/cufft}
\showURL{%
\tempurl}


\bibitem[NVIDIA(2021c)]%
        {curand}
\bibfield{author}{\bibinfo{person}{NVIDIA}.} \bibinfo{year}{2021}\natexlab{c}.
\newblock \bibinfo{title}{cuRAND: Random number generation on NVIDIA GPUs}.
\newblock
\urldef\tempurl%
\url{https://developer.nvidia.com/curand}
\showURL{%
\tempurl}


\bibitem[NVIDIA(2021d)]%
        {cusolver}
\bibfield{author}{\bibinfo{person}{NVIDIA}.} \bibinfo{year}{2021}\natexlab{d}.
\newblock \bibinfo{title}{cuSOLVER: Direct linear solvers on NVIDIA GPUs}.
\newblock
\urldef\tempurl%
\url{https://developer.nvidia.com/cusolver}
\showURL{%
\tempurl}


\bibitem[NVIDIA(2021e)]%
        {cusparse}
\bibfield{author}{\bibinfo{person}{NVIDIA}.} \bibinfo{year}{2021}\natexlab{e}.
\newblock \bibinfo{title}{cuSPARSE: GPU library APIs for sparse computation}.
\newblock
\urldef\tempurl%
\url{https://developer.nvidia.com/cusparse}
\showURL{%
\tempurl}


\bibitem[NVIDIA(2021f)]%
        {cutensor}
\bibfield{author}{\bibinfo{person}{NVIDIA}.} \bibinfo{year}{2021}\natexlab{f}.
\newblock \bibinfo{title}{cuTENSOR: A high-performance CUDA library for tensor
  primitives}.
\newblock
\urldef\tempurl%
\url{https://docs.nvidia.com/cuda/cutensor/latest/index.html}
\showURL{%
\tempurl}


\bibitem[NVIDIA(2021g)]%
        {matx}
\bibfield{author}{\bibinfo{person}{NVIDIA}.} \bibinfo{year}{2021}\natexlab{g}.
\newblock \bibinfo{title}{MATX: GPU-accelerated numerical computing in modern
  C++}.
\newblock
\urldef\tempurl%
\url{https://github.com/NVIDIA/MatX}
\showURL{%
\tempurl}


\bibitem[NVIDIA(2021h)]%
        {castVisual}
\bibfield{author}{\bibinfo{person}{NVIDIA}.} \bibinfo{year}{2021}\natexlab{h}.
\newblock \bibinfo{title}{RAPIDS AI cuDF: [Bug] Cast from \texttt{decimal64} to
  \texttt{decimal128} is really slow (issue \#9597)}.
\newblock
\urldef\tempurl%
\url{https://github.com/rapidsai/cudf/issues/9597}
\showURL{%
\tempurl}


\bibitem[NVIDIA(2021i)]%
        {datatype}
\bibfield{author}{\bibinfo{person}{NVIDIA}.} \bibinfo{year}{2021}\natexlab{i}.
\newblock \bibinfo{title}{RAPIDS AI cuDF: [Bug] CUDA error invalid value when
  creating \texttt{cudf.series} from float16 CuPy series (issue \#9065)}.
\newblock
\urldef\tempurl%
\url{https://github.com/rapidsai/cudf/issues/9065}
\showURL{%
\tempurl}


\bibitem[NVIDIA(2021j)]%
        {casting}
\bibfield{author}{\bibinfo{person}{NVIDIA}.} \bibinfo{year}{2021}\natexlab{j}.
\newblock \bibinfo{title}{RAPIDS AI cuDF: [Bug] \texttt{libcudf} unary cast
  returns unexpected results when casting between decimal and int types (issue
  \#7689)}.
\newblock
\urldef\tempurl%
\url{https://github.com/rapidsai/cudf/issues/7689}
\showURL{%
\tempurl}


\bibitem[NVIDIA(2023a)]%
        {cudaprogrammingguide}
\bibfield{author}{\bibinfo{person}{NVIDIA}.} \bibinfo{year}{2023}\natexlab{a}.
\newblock \bibinfo{title}{CUDA C++ programming guide}.
\newblock
\urldef\tempurl%
\url{https://docs.nvidia.com/cuda/cuda-c-programming-guide/index.html#floating-point-standard}
\showURL{%
\tempurl}


\bibitem[NVIDIA(2023b)]%
        {Thrust}
\bibfield{author}{\bibinfo{person}{NVIDIA}.} \bibinfo{year}{2023}\natexlab{b}.
\newblock \bibinfo{title}{Thrust: The C++ parallel algorithms library}.
\newblock
\urldef\tempurl%
\url{https://docs.nvidia.com/cuda/thrust/index.html}
\showURL{%
\tempurl}


\bibitem[NVIDIA(2025)]%
        {nvidiagpus}
\bibfield{author}{\bibinfo{person}{NVIDIA}.} \bibinfo{year}{2025}\natexlab{}.
\newblock \bibinfo{title}{NVIDIA graphics cards}.
\newblock
\urldef\tempurl%
\url{https://www.nvidia.com/en-us/geforce/graphics-cards/}
\showURL{%
\tempurl}


\bibitem[Pham et~al\mbox{.}(2019)]%
        {DLGPU4}
\bibfield{author}{\bibinfo{person}{Hung~Viet Pham}, \bibinfo{person}{Thibaud
  Lutellier}, \bibinfo{person}{Weizhen Qi}, {and} \bibinfo{person}{Lin Tan}.}
  \bibinfo{year}{2019}\natexlab{}.
\newblock \showarticletitle{CRADLE: Cross-backend validation to detect and
  localize bugs in deep learning libraries}. In
  \bibinfo{booktitle}{\emph{Proceedings of the 41st IEEE/ACM International
  Conference on Software Engineering (ICSE)}}. \bibinfo{pages}{1027--1038}.
\newblock
\href{https://doi.org/10.1109/ICSE.2019.00107}{doi:\nolinkurl{10.1109/ICSE.2019.00107}}


\bibitem[Polok and Smrz(2016)]%
        {polok2016increasing}
\bibfield{author}{\bibinfo{person}{Lukas Polok} {and} \bibinfo{person}{Pavel
  Smrz}.} \bibinfo{year}{2016}\natexlab{}.
\newblock \showarticletitle{Increasing double precision throughput on NVIDIA
  Maxwell GPUs}. In \bibinfo{booktitle}{\emph{Proceedings of the 24th High
  Performance Computing Symposium (HPC)}}. \bibinfo{pages}{1--8}.
\newblock
\href{https://doi.org/10.22360/SpringSim.2016.HPC.032}{doi:\nolinkurl{10.22360/SpringSim.2016.HPC.032}}


\bibitem[Research(2022)]%
        {special2}
\bibfield{author}{\bibinfo{person}{Meta Research}.}
  \bibinfo{year}{2022}\natexlab{}.
\newblock \bibinfo{title}{PyTorch: \texttt{torch.norm} gives NaN gradient when
  I input small-value float16 tensor (issue \#43211)}.
\newblock
\urldef\tempurl%
\url{https://github.com/pytorch/pytorch/issues/43211}
\showURL{%
\tempurl}


\bibitem[Shen et~al\mbox{.}(2021)]%
        {DLGPU6}
\bibfield{author}{\bibinfo{person}{Qingchao Shen}, \bibinfo{person}{Haoyang
  Ma}, \bibinfo{person}{Junjie Chen}, \bibinfo{person}{Yongqiang Tian},
  \bibinfo{person}{Shing-Chi Cheung}, {and} \bibinfo{person}{Xiang Chen}.}
  \bibinfo{year}{2021}\natexlab{}.
\newblock \showarticletitle{A comprehensive study of deep learning compiler
  bugs}. In \bibinfo{booktitle}{\emph{Proceedings of the 29th ACM Joint Meeting
  on European Software Engineering Conference and Symposium on the Foundations
  of Software Engineering (ESEC/FSE)}}. \bibinfo{pages}{968--980}.
\newblock
\href{https://doi.org/10.1145/3468264.3468591}{doi:\nolinkurl{10.1145/3468264.3468591}}


\bibitem[Solt et~al\mbox{.}(2024)]%
        {solt2024cascade}
\bibfield{author}{\bibinfo{person}{Flavien Solt}, \bibinfo{person}{Katharina
  Ceesay-Seitz}, {and} \bibinfo{person}{Kaveh Razavi}.}
  \bibinfo{year}{2024}\natexlab{}.
\newblock \showarticletitle{Cascade: CPU fuzzing via intricate program
  generation}. In \bibinfo{booktitle}{\emph{Proceedings of the 33rd USENIX
  Security Symposium (USENIX Security)}}. \bibinfo{pages}{5341--5358}.
\newblock
\showISBNx{978-1-939133-44-1}


\bibitem[Sorensen and Donaldson(2016)]%
        {sorensen2016exposing}
\bibfield{author}{\bibinfo{person}{Tyler Sorensen} {and}
  \bibinfo{person}{Alastair~F. Donaldson}.} \bibinfo{year}{2016}\natexlab{}.
\newblock \showarticletitle{Exposing errors related to weak memory in GPU
  applications}. In \bibinfo{booktitle}{\emph{Proceedings of the 37th ACM
  SIGPLAN Conference on Programming Language Design and Implementation
  (PLDI)}}. \bibinfo{pages}{100--113}.
\newblock
\href{https://doi.org/10.1145/2908080.2908114}{doi:\nolinkurl{10.1145/2908080.2908114}}


\bibitem[Surineni et~al\mbox{.}(2017)]%
        {surineni2017understanding}
\bibfield{author}{\bibinfo{person}{Sruthikesh Surineni},
  \bibinfo{person}{Ruidong Gu}, \bibinfo{person}{Huyen Nguyen}, {and}
  \bibinfo{person}{Michela Becchi}.} \bibinfo{year}{2017}\natexlab{}.
\newblock \showarticletitle{Understanding the performance-accuracy tradeoffs of
  floating-point arithmetic on GPUs}. In \bibinfo{booktitle}{\emph{Proceedings
  of the IEEE International Symposium on Workload Characterization (IISWC)}}.
  \bibinfo{pages}{207--218}.
\newblock
\href{https://doi.org/10.1109/IISWC.2017.8167778}{doi:\nolinkurl{10.1109/IISWC.2017.8167778}}


\bibitem[Tang et~al\mbox{.}(2010)]%
        {tang2010perturbing}
\bibfield{author}{\bibinfo{person}{Enyi Tang}, \bibinfo{person}{Earl Barr},
  \bibinfo{person}{Xuandong Li}, {and} \bibinfo{person}{Zhendong Su}.}
  \bibinfo{year}{2010}\natexlab{}.
\newblock \showarticletitle{Perturbing numerical calculations for statistical
  analysis of floating-point program (in)stability}. In
  \bibinfo{booktitle}{\emph{Proceedings of the 19th ACM SIGSOFT International
  Symposium on Software Testing and Analysis (ISSTA)}}.
  \bibinfo{pages}{131--142}.
\newblock
\href{https://doi.org/10.1145/1831708.1831724}{doi:\nolinkurl{10.1145/1831708.1831724}}


\bibitem[Team(2021a)]%
        {cupy}
\bibfield{author}{\bibinfo{person}{CuPy~Development Team}.}
  \bibinfo{year}{2021}\natexlab{a}.
\newblock \bibinfo{title}{CuPy: NumPy and SciPy for GPU}.
\newblock
\urldef\tempurl%
\url{https://github.com/cupy/cupy}
\showURL{%
\tempurl}


\bibitem[Team(2021b)]%
        {correctness1}
\bibfield{author}{\bibinfo{person}{CuPy~Development Team}.}
  \bibinfo{year}{2021}\natexlab{b}.
\newblock \bibinfo{title}{CuPy: Passing CuPy functions into
  \texttt{cupyx.scipy.ndimage.filters.generic\_filter} causes a TypeError
  (issue \#3909)}.
\newblock
\urldef\tempurl%
\url{https://github.com/cupy/cupy/issues/3909}
\showURL{%
\tempurl}


\bibitem[Team(2021c)]%
        {environment1}
\bibfield{author}{\bibinfo{person}{CuPy~Development Team}.}
  \bibinfo{year}{2021}\natexlab{c}.
\newblock \bibinfo{title}{CuPy: \texttt{cp.matmul} slower than
  \texttt{np.matmul} (issue \#4891)}.
\newblock
\urldef\tempurl%
\url{https://github.com/cupy/cupy/issues/4891}
\showURL{%
\tempurl}


\bibitem[Team(2021d)]%
        {Accuracy1}
\bibfield{author}{\bibinfo{person}{CuPy~Development Team}.}
  \bibinfo{year}{2021}\natexlab{d}.
\newblock \bibinfo{title}{CuPy: \texttt{ndimage.map\_coordinates} performs
  incorrectly with several order values (issue \#4550)}.
\newblock
\urldef\tempurl%
\url{https://github.com/cupy/cupy/issues/4550}
\showURL{%
\tempurl}


\bibitem[Team(2021e)]%
        {Accuracy2}
\bibfield{author}{\bibinfo{person}{CuPy~Development Team}.}
  \bibinfo{year}{2021}\natexlab{e}.
\newblock \bibinfo{title}{cupyx.scipy.signal.convolve2d: Integer overflow when
  using mixed dtypes (issue \#6047)}.
\newblock
\urldef\tempurl%
\url{https://github.com/cupy/cupy/issues/6047}
\showURL{%
\tempurl}


\bibitem[Team(2022a)]%
        {CupyNaN}
\bibfield{author}{\bibinfo{person}{CuPy~Development Team}.}
  \bibinfo{year}{2022}\natexlab{a}.
\newblock \bibinfo{title}{CuPy: Unexpected NaN when using big-endian arrays
  (issue \#3652)}.
\newblock
\urldef\tempurl%
\url{https://github.com/cupy/cupy/issues/3652}
\showURL{%
\tempurl}


\bibitem[Team(2022b)]%
        {user-defined}
\bibfield{author}{\bibinfo{person}{CuPy~Development Team}.}
  \bibinfo{year}{2022}\natexlab{b}.
\newblock \bibinfo{title}{CuPy: User-defined kernels}.
\newblock
\urldef\tempurl%
\url{https://docs.cupy.dev/en/stable/user_guide/kernel.html}
\showURL{%
\tempurl}


\bibitem[Team(2021f)]%
        {numba}
\bibfield{author}{\bibinfo{person}{Numba~Development Team}.}
  \bibinfo{year}{2021}\natexlab{f}.
\newblock \bibinfo{title}{Numba: NumPy aware dynamic Python compiler using
  LLVM}.
\newblock
\urldef\tempurl%
\url{https://github.com/numba/numba}
\showURL{%
\tempurl}


\bibitem[Tiwari et~al\mbox{.}(2015)]%
        {tiwari2015understanding}
\bibfield{author}{\bibinfo{person}{Devesh Tiwari}, \bibinfo{person}{Saurabh
  Gupta}, \bibinfo{person}{James Rogers}, \bibinfo{person}{Don Maxwell},
  \bibinfo{person}{Paolo Rech}, \bibinfo{person}{Sudharshan Vazhkudai},
  \bibinfo{person}{Daniel Oliveira}, \bibinfo{person}{Dave Londo},
  \bibinfo{person}{Nathan DeBardeleben}, \bibinfo{person}{Philippe Navaux},
  {et~al\mbox{.}}} \bibinfo{year}{2015}\natexlab{}.
\newblock \showarticletitle{Understanding GPU errors on large-scale HPC systems
  and the implications for system design and operation}. In
  \bibinfo{booktitle}{\emph{Proceedings of the 21st IEEE International
  Symposium on High Performance Computer Architecture (HPCA)}}.
  \bibinfo{pages}{331--342}.
\newblock
\href{https://doi.org/10.1109/HPCA.2015.7056044}{doi:\nolinkurl{10.1109/HPCA.2015.7056044}}


\bibitem[Vedantu(2022)]%
        {operatordef}
\bibfield{author}{\bibinfo{person}{Vedantu}.} \bibinfo{year}{2022}\natexlab{}.
\newblock \bibinfo{title}{Numerical expression}.
\newblock
\urldef\tempurl%
\url{https://www.vedantu.com/maths/numerical-expression}
\showURL{%
\tempurl}


\bibitem[Wang et~al\mbox{.}(2022)]%
        {DLGPU3}
\bibfield{author}{\bibinfo{person}{Gan Wang}, \bibinfo{person}{Zan Wang},
  \bibinfo{person}{Junjie Chen}, \bibinfo{person}{Xiang Chen}, {and}
  \bibinfo{person}{Ming Yan}.} \bibinfo{year}{2022}\natexlab{}.
\newblock \showarticletitle{An empirical study on numerical bugs in deep
  learning programs}. In \bibinfo{booktitle}{\emph{Proceedings of the 37th
  IEEE/ACM International Conference on Automated Software Engineering (ASE)}}.
  \bibinfo{pages}{1--5}.
\newblock
\href{https://doi.org/10.1145/3551349.3559561}{doi:\nolinkurl{10.1145/3551349.3559561}}


\bibitem[Wang et~al\mbox{.}(2016)]%
        {wang2016detecting}
\bibfield{author}{\bibinfo{person}{Ran Wang}, \bibinfo{person}{Daming Zou},
  \bibinfo{person}{Xinrui He}, \bibinfo{person}{Yingfei Xiong},
  \bibinfo{person}{Lu Zhang}, {and} \bibinfo{person}{Gang Huang}.}
  \bibinfo{year}{2016}\natexlab{}.
\newblock \showarticletitle{Detecting and fixing precision-specific operations
  for measuring floating-point errors}. In
  \bibinfo{booktitle}{\emph{Proceedings of the 24th ACM SIGSOFT International
  Symposium on Foundations of Software Engineering (FSE)}}.
  \bibinfo{pages}{619--630}.
\newblock
\href{https://doi.org/10.1145/2950290.2950355}{doi:\nolinkurl{10.1145/2950290.2950355}}


\bibitem[Whitehead and Fit-Florea(2011)]%
        {whitehead2011precision}
\bibfield{author}{\bibinfo{person}{Nathan Whitehead} {and}
  \bibinfo{person}{Alex Fit-Florea}.} \bibinfo{year}{2011}\natexlab{}.
\newblock \showarticletitle{Precision \& performance: Floating point and IEEE
  754 compliance for NVIDIA GPUs}.
\newblock \bibinfo{journal}{\emph{NVIDIA Documentation}}  \bibinfo{volume}{21}
  (\bibinfo{year}{2011}), \bibinfo{pages}{18749--19424}.
\newblock


\bibitem[Wilt(2013a)]%
        {cudahandbook}
\bibfield{author}{\bibinfo{person}{Nicholas Wilt}.}
  \bibinfo{year}{2013}\natexlab{a}.
\newblock \bibinfo{booktitle}{\emph{The CUDA Handbook: A Comprehensive Guide to
  GPU Programming}}.
\newblock \bibinfo{publisher}{Pearson Education}.
\newblock


\bibitem[Wilt(2013b)]%
        {admin_2013}
\bibfield{author}{\bibinfo{person}{Nicholas Wilt}.}
  \bibinfo{year}{2013}\natexlab{b}.
\newblock \bibinfo{title}{Floating point: CPU and GPU differences}.
\newblock
\newblock
\shownote{\url{http://www.cudahandbook.com/2013/08/floating-point-cpu-and-gpu-differences/}}.


\bibitem[Wu et~al\mbox{.}(2020)]%
        {wu2020simulee}
\bibfield{author}{\bibinfo{person}{Mingyuan Wu}, \bibinfo{person}{Yicheng
  Ouyang}, \bibinfo{person}{Husheng Zhou}, \bibinfo{person}{Lingming Zhang},
  \bibinfo{person}{Cong Liu}, {and} \bibinfo{person}{Yuqun Zhang}.}
  \bibinfo{year}{2020}\natexlab{}.
\newblock \showarticletitle{Simulee: Detecting CUDA synchronization bugs via
  memory-access modeling}. In \bibinfo{booktitle}{\emph{Proceedings of the 42nd
  IEEE/ACM International Conference on Software Engineering (ICSE)}}.
  \bibinfo{pages}{937--948}.
\newblock
\href{https://doi.org/10.1145/3377811.3380358}{doi:\nolinkurl{10.1145/3377811.3380358}}


\bibitem[Xiao et~al\mbox{.}(2022)]%
        {DLGPU1}
\bibfield{author}{\bibinfo{person}{Dongwei Xiao}, \bibinfo{person}{Zhibo Liu},
  \bibinfo{person}{Yuanyuan Yuan}, \bibinfo{person}{Qi Pang}, {and}
  \bibinfo{person}{Shuai Wang}.} \bibinfo{year}{2022}\natexlab{}.
\newblock \showarticletitle{Metamorphic testing of deep learning compilers}.
\newblock \bibinfo{journal}{\emph{Proceedings of the ACM on Measurement and
  Analysis of Computing Systems (POMACS)}}  \bibinfo{volume}{6}
  (\bibinfo{year}{2022}), \bibinfo{pages}{1--28}.
\newblock
\href{https://doi.org/10.1145/3508035}{doi:\nolinkurl{10.1145/3508035}}


\bibitem[Yan et~al\mbox{.}(2021)]%
        {DLbug}
\bibfield{author}{\bibinfo{person}{Ming Yan}, \bibinfo{person}{Junjie Chen},
  \bibinfo{person}{Xiangyu Zhang}, \bibinfo{person}{Lin Tan},
  \bibinfo{person}{Gan Wang}, {and} \bibinfo{person}{Zan Wang}.}
  \bibinfo{year}{2021}\natexlab{}.
\newblock \showarticletitle{Exposing numerical bugs in deep learning via
  gradient back-propagation}. In \bibinfo{booktitle}{\emph{Proceedings of the
  29th ACM Joint Meeting on European Software Engineering Conference and
  Symposium on the Foundations of Software Engineering (ESEC/FSE)}}.
  \bibinfo{pages}{627--638}.
\newblock
\href{https://doi.org/10.1145/3468264.3468612}{doi:\nolinkurl{10.1145/3468264.3468612}}


\bibitem[Yang et~al\mbox{.}(2012)]%
        {yang2012fixing111}
\bibfield{author}{\bibinfo{person}{Yi Yang}, \bibinfo{person}{Ping Xiang},
  \bibinfo{person}{Mike Mantor}, {and} \bibinfo{person}{Huiyang Zhou}.}
  \bibinfo{year}{2012}\natexlab{}.
\newblock \showarticletitle{{Fixing performance bugs: An empirical study of
  open-source GPGPU programs}}. In \bibinfo{booktitle}{\emph{Proceedings of the
  41st IEEE International Conference on Parallel Processing (ICPP)}}.
  \bibinfo{pages}{329--339}.
\newblock
\href{https://doi.org/10.1109/ICPP.2012.30}{doi:\nolinkurl{10.1109/ICPP.2012.30}}


\bibitem[Yi et~al\mbox{.}(2019)]%
        {yi2019efficient}
\bibfield{author}{\bibinfo{person}{Xin Yi}, \bibinfo{person}{Liqian Chen},
  \bibinfo{person}{Xiaoguang Mao}, {and} \bibinfo{person}{Tao Ji}.}
  \bibinfo{year}{2019}\natexlab{}.
\newblock \showarticletitle{Efficient automated repair of high floating-point
  errors in numerical libraries}.
\newblock \bibinfo{journal}{\emph{Proceedings of the ACM on Programming
  Languages (POPL)}}  \bibinfo{volume}{3} (\bibinfo{year}{2019}),
  \bibinfo{pages}{1--29}.
\newblock
\href{https://doi.org/10.1145/3290369}{doi:\nolinkurl{10.1145/3290369}}


\bibitem[Zeller et~al\mbox{.}(2023)]%
        {greybox}
\bibfield{author}{\bibinfo{person}{Andreas Zeller}, \bibinfo{person}{Rahul
  Gopinath}, \bibinfo{person}{Marcel B{\"o}hme}, \bibinfo{person}{Gordon
  Fraser}, {and} \bibinfo{person}{Christian Holler}.}
  \bibinfo{year}{2023}\natexlab{}.
\newblock \showarticletitle{Greybox Fuzzing}.
\newblock In \bibinfo{booktitle}{\emph{The Fuzzing Book}}.
  \bibinfo{publisher}{CISPA Helmholtz Center for Information Security}.
\newblock
\urldef\tempurl%
\url{https://www.fuzzingbook.org/html/GreyboxFuzzer.html}
\showURL{%
\tempurl}


\bibitem[Zhang et~al\mbox{.}(2020)]%
        {DLGPU2}
\bibfield{author}{\bibinfo{person}{Yuhao Zhang}, \bibinfo{person}{Luyao Ren},
  \bibinfo{person}{Liqian Chen}, \bibinfo{person}{Yingfei Xiong},
  \bibinfo{person}{Shing-Chi Cheung}, {and} \bibinfo{person}{Tao Xie}.}
  \bibinfo{year}{2020}\natexlab{}.
\newblock \showarticletitle{{Detecting numerical bugs in neural network
  architectures}}. In \bibinfo{booktitle}{\emph{Proceedings of the 28th ACM
  Joint Meeting on European Software Engineering Conference and Symposium on
  the Foundations of Software Engineering (ESEC/FSE)}}.
  \bibinfo{pages}{826--837}.
\newblock
\href{https://doi.org/10.1145/3368089.3409720}{doi:\nolinkurl{10.1145/3368089.3409720}}


\bibitem[Zou et~al\mbox{.}(2015)]%
        {zou2015genetic}
\bibfield{author}{\bibinfo{person}{Daming Zou}, \bibinfo{person}{Ran Wang},
  \bibinfo{person}{Yingfei Xiong}, \bibinfo{person}{Lu Zhang},
  \bibinfo{person}{Zhendong Su}, {and} \bibinfo{person}{Hong Mei}.}
  \bibinfo{year}{2015}\natexlab{}.
\newblock \showarticletitle{{A genetic algorithm for detecting significant
  floating-point inaccuracies}}. In \bibinfo{booktitle}{\emph{Proceedings of
  the 37th IEEE/ACM International Conference on Software Engineering (ICSE)}},
  Vol.~\bibinfo{volume}{1}. \bibinfo{pages}{529--539}.
\newblock
\href{https://doi.org/10.1109/ICSE.2015.70}{doi:\nolinkurl{10.1109/ICSE.2015.70}}


\bibitem[Zou et~al\mbox{.}(2019)]%
        {zou2019detecting}
\bibfield{author}{\bibinfo{person}{Daming Zou}, \bibinfo{person}{Muhan Zeng},
  \bibinfo{person}{Yingfei Xiong}, \bibinfo{person}{Zhoulai Fu},
  \bibinfo{person}{Lu Zhang}, {and} \bibinfo{person}{Zhendong Su}.}
  \bibinfo{year}{2019}\natexlab{}.
\newblock \showarticletitle{Detecting floating-point errors via atomic
  conditions}.
\newblock \bibinfo{journal}{\emph{Proceedings of the ACM on Programming
  Languages (POPL)}}  \bibinfo{volume}{4} (\bibinfo{year}{2019}),
  \bibinfo{pages}{1--27}.
\newblock
\href{https://doi.org/10.1145/3371128}{doi:\nolinkurl{10.1145/3371128}}


\end{thebibliography}
